\RequirePackage{fix-cm}
\documentclass[twocolumn,epjc3]{svjour3}  
\smartqed  
\RequirePackage{graphicx}
\RequirePackage{mathptmx}      
\RequirePackage{flushend}
\RequirePackage[numbers,sort&compress]{natbib}
\RequirePackage[colorlinks,citecolor=blue,urlcolor=blue,linkcolor=blue]{hyperref}
\usepackage{revsymb}
\usepackage{lipsum}
\usepackage{amsmath} 
\usepackage{verbatim} 
\usepackage{color} 
\usepackage{subfigure} 
\usepackage{hyperref} 
\usepackage{dcolumn}
\usepackage{epsfig}
\usepackage{tabularx}
\usepackage{amssymb}
\usepackage{amsmath} 
\usepackage{afterpage} 
\usepackage{float}
\usepackage{sidecap}
\usepackage{placeins}
\newcommand{\secref}[1]{section~\ref{#1}}
\usepackage{stfloats}
\usepackage{lipsum}

\journalname{EPJA}

\begin{document}

\title{Charged-pion production in $\mathbf{Au+Au}$ collisions at $\mathbf{\sqrt{s_{NN}}}$ = 2.4$\mathbf{~GeV}$}

\author{HADES collaboration \\[5bp]
J.~Adamczewski-Musch$^{5}$, O.~Arnold$^{11,10}$, C.~Behnke$^{9}$, A.~Belounnas$^{17}$,
A.~Belyaev$^{8}$, J.C.~Berger-Chen$^{11,10}$, A.~Blanco$^{2}$, C.~Blume$^{9}$, M.~B\"{o}hmer$^{11}$,
P.~Bordalo$^{2}$
% ...
}

\institute{
\mbox{} \\[-8bp]
\mbox{$^{1}$Institute of Physics, Slovak Academy of Sciences, 84228~Bratislava, Slovakia}\\
\mbox{$^{2}$LIP-Laborat\'{o}rio de Instrumenta\c{c}\~{a}o e F\'{\i}sica Experimental de Part\'{\i}culas , 3004-516~Coimbra, Portugal}\\
\mbox{$^{3}$Institute of Nuclear Physics, Polish Academy of Sciences, 31342~Krak\'{o}w, Poland}\\
\mbox{$^{4}$Smoluchowski Institute of Physics, Jagiellonian University of Cracow, 30-059~Krak\'{o}w, Poland}\\
\mbox{$^{5}$GSI Helmholtzzentrum f\"{u}r Schwerionenforschung GmbH, 64291~Darmstadt, Germany}\\
\mbox{$^{6}$Technische Universit\"{a}t Darmstadt, 64289~Darmstadt, Germany}\\
\mbox{$^{7}$Institut f\"{u}r Strahlenphysik, Helmholtz-Zentrum Dresden-Rossendorf, 01314~Dresden, Germany}\\
\mbox{$^{8}$Joint Institute of Nuclear Research, 141980~Dubna, Russia}\\
\mbox{$^{9}$Institut f\"{u}r Kernphysik, Goethe-Universit\"{a}t, 60438 ~Frankfurt, Germany}\\
\mbox{$^{10}$Excellence Cluster 'Origin and Structure of the Universe' , 85748~Garching, Germany}\\
\mbox{$^{11}$Physik Department E62, Technische Universit\"{a}t M\"{u}nchen, 85748~Garching, Germany}\\
\mbox{$^{12}$II.Physikalisches Institut, Justus Liebig Universit\"{a}t Giessen, 35392~Giessen, Germany}\\
\mbox{$^{13}$Institute for Nuclear Research, Russian Academy of Science, 117312~Moscow, Russia}\\
\mbox{$^{14}$Institute of Theoretical and Experimental Physics, 117218~Moscow, Russia}\\
\mbox{$^{15}$National Research Nuclear University MEPhI (Moscow Engineering Physics Institute), 115409~Moscow, Russia}\\
\mbox{$^{16}$Department of Physics, University of Cyprus, 1678~Nicosia, Cyprus}\\
\mbox{$^{17}$Laboratoire de Physique des 2 infinis Irène Joliot-Curie, Université Paris-Saclay, CNRS-IN2P3. , F-91405~Orsay , France}\\
\mbox{$^{18}$Nuclear Physics Institute, The Czech Academy of Sciences, 25068~Rez, Czech Republic}\\
\mbox{$^{19}$LabCAF. F. F\'{\i}sica, Univ. de Santiago de Compostela, 15706~Santiago de Compostela, Spain}\\
\mbox{$^{20}$Uniwersytet Warszawski - Instytut Fizyki Do\'{s}wiadczalnej, 02-093~Warszawa, Poland}\\
\mbox{$^{21}$Dipartimento di Fisica and INFN, Universit\`{a} di Torino, 10125~Torino, Italy}\\
	\\
\mbox{$^{a}$ also at Coimbra Polytechnic - ISEC, ~Coimbra, Portugal}\\
\mbox{$^{b}$ also at Technische Universit\"{a}t Dresden, 01062~Dresden, Germany}\\
\mbox{$^{c}$ also at Frederick University, 1036~Nicosia, Cyprus}\\
\mbox{$^{\dag}$ deceased} \\
\mbox{ e-mail: hades-info@gsi.de}\\
}

\date{\today}

\authorrunning{J.~Adamczewski-Musch et al.}
\titlerunning{Charged pion production in $\mathbf{Au+Au}$ collisions at $\mathbf{\sqrt{s_{NN}}}$ = 2.4 $\mathbf{GeV}$}

\maketitle

\abstract{
We present high-statistic data on charged-pion emission from Au+Au collisions at
$\sqrt{s_{\rm{NN}}}$ = 2.4~GeV (corresponding to $E_{beam}$ = 1.23~A~GeV) in four centrality classes
in the range 0 - 40$\%$ of the most central collisions. The data are analyzed as a function of transverse momentum, 
transverse mass, rapidity, and polar angle. 
Pion multiplicity per participating nucleon decreases moderately with increasing centrality.
The polar angular distributions are found to be non-isotropic even for the most central event class.
Our results on pion multiplicity fit well into the general trend of the world data, but undershoot by $2.5~\sigma$ data from
the FOPI experiment measured at slightly lower beam energy. 
We compare our data to state-of-the-art transport model calculations (PHSD, IQMD, PHQMD, GiBUU and SMASH) and find
substantial differences between the measurement and the results of these calculations. 
}

\section{Introduction}
\label{sec1}
Heavy-ion collisions at relativistic energies allow the study of bulk properties of strongly interacting matter at high temperatures and densities. 
In these studies, the phase-space distributions of various final-state particles are analyzed and compared to the corresponding
distributions in nucleon-nucleon interactions in order to disentangle bulk phenomena from the trivial superposition of elementary interactions.
The particles present in the final state of relativistic nuclear collisions carry information about the initial state, e.g. the impact parameter,
about the properties of the high-density phase of the system, e.g. the pressure and its gradients, and about the expansion and freeze-out conditions
of the produced strongly-interacting matter, often called fireball.  The final-state particles are either surviving nucleons, nuclear clusters, or newly-produced particles. 
Being the lightest mesons, pions are the pseudo-Goldstone bosons of SU2 reflecting the approximate spontaneous breaking of chiral symmetry of the QCD.
Hence, they are a measure for the degree of excitation in a gas of hadrons \cite{Stock:1985xe}.  They have an isospin of one and come in the three charge states
$\pi^{+}$, $\pi^{-}$ and $\pi^{0}$, and they are the only abundantly produced particles in the few-GeV energy range.
Their yields, phase-space distributions, and multi-particle correlations carry information about all stages of the collision.  
\par
In this paper, we present experimental data on charged-pion production in centrality-selected Au+Au collisions at $\sqrt{s_{NN}}$=2.4~GeV
(corresponding to a beam kinetic energy of $E_{beam}$ = 1.23~A~GeV on fixed target).  Our results profit from high statistics and thus complement
and extend earlier studies of pion production at similar energies and with heavy nuclei~\cite{Muntz:1995am,Wagner:1997sa,Reisdorf:2006ie,Pelte:1997rg,Wolf:1998vn,Averbeck:2000sn,Averbeck:1997ma,Holzmann:1997mu,Harris:1987md,Klay:2003zf,Nagamiya:1981sd}.
They cover rapidity and transverse-momentum (or mass) distributions, as well as derived quantities.  Some of the latter are analyzed as a function of the collision centrality.
Special emphasis is put on the comparison of the experimental findings with results from microscopic model calculations.  Detailed investigations of spectra generated in thermal
models in comparison to experimental data are also ongoing and will be discussed in future publications.  First results on two-pion correlations have recently
been published~\cite{Adamczewski-Musch:2018owj,Adamczewski-Musch:2019dlf}.   Multi-pion correlation and collective-flow studies are the subject of separate forthcoming articles. 
\par
After describing the experimental setup and the analysis methods in \secref{sec2}, we present transverse-momentum ($p_{t}$) and reduced transverse-mass spectra ($m_{t} - m_{0}$),
as well as rapidity distributions which are used to determine the multiplicities of charged pions in \secref{sec3}.  In~\secref{sec3_1}, our result on the pion yield is compared
to the pion excitation function for beam kinetic energies ranging from threshold up to 10~$A$~GeV.  It is well established that pions at SIS18 energies are emitted preferably
in the forward/backward direction.  In section \secref{sec3_2}, we present the centrality and momentum dependence of the parameter $A_2$ which quantifies the pion anisotropy.
The presentation of our results ends with a detailed comparison of the observables discussed in the previous sections to five state-of-the-art microscopic
models \cite{Hartnack:1997ez,Cassing:1999es,Buss:2011mx,Petersen:2018jag,Weil:2016zrk,Aichelin:2019tnk} in \secref{sec4}.
\par
We note in passing that the measured pion yields are important for the normalization of the dielectron data obtained in the same experiment~\cite{Adamczewski-Musch:2019byl}.
The low-energy region of the invariant-mass spectra of dielectrons is dominated by the decay products of neutral pions.  The charged pions can be used to construct triple differential
momentum distributions of neutral pions and their decays which allow to constrain the part of the dielectron spectrum originating from pion decays. 

\FloatBarrier 
\section{Experimental setup and data analysis}
\label{sec2}
\begin{figure*}
	\centering
	\subfigure[]{\includegraphics[scale=0.4]{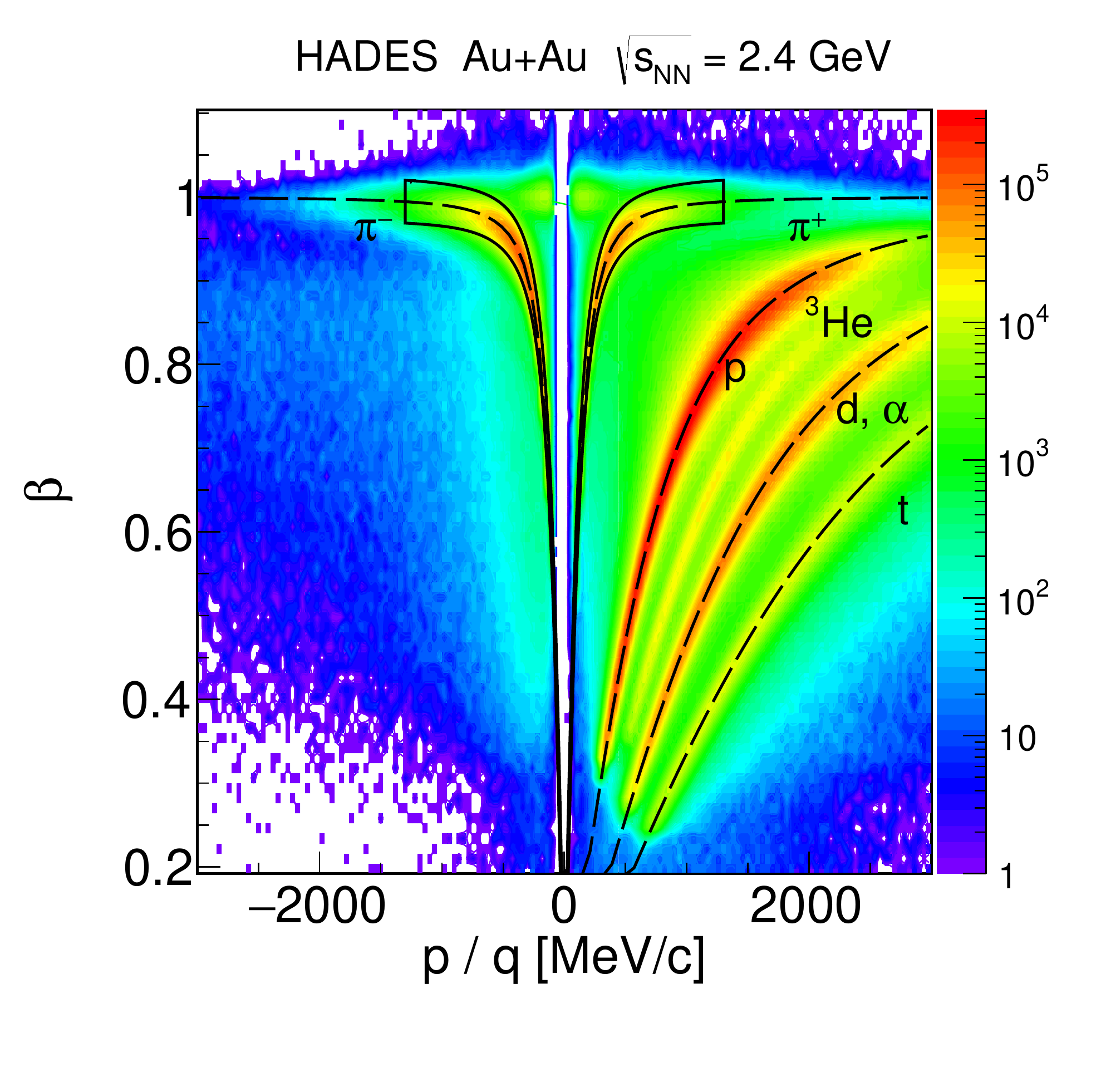}}\quad
	\subfigure[]{\includegraphics[scale=0.4]{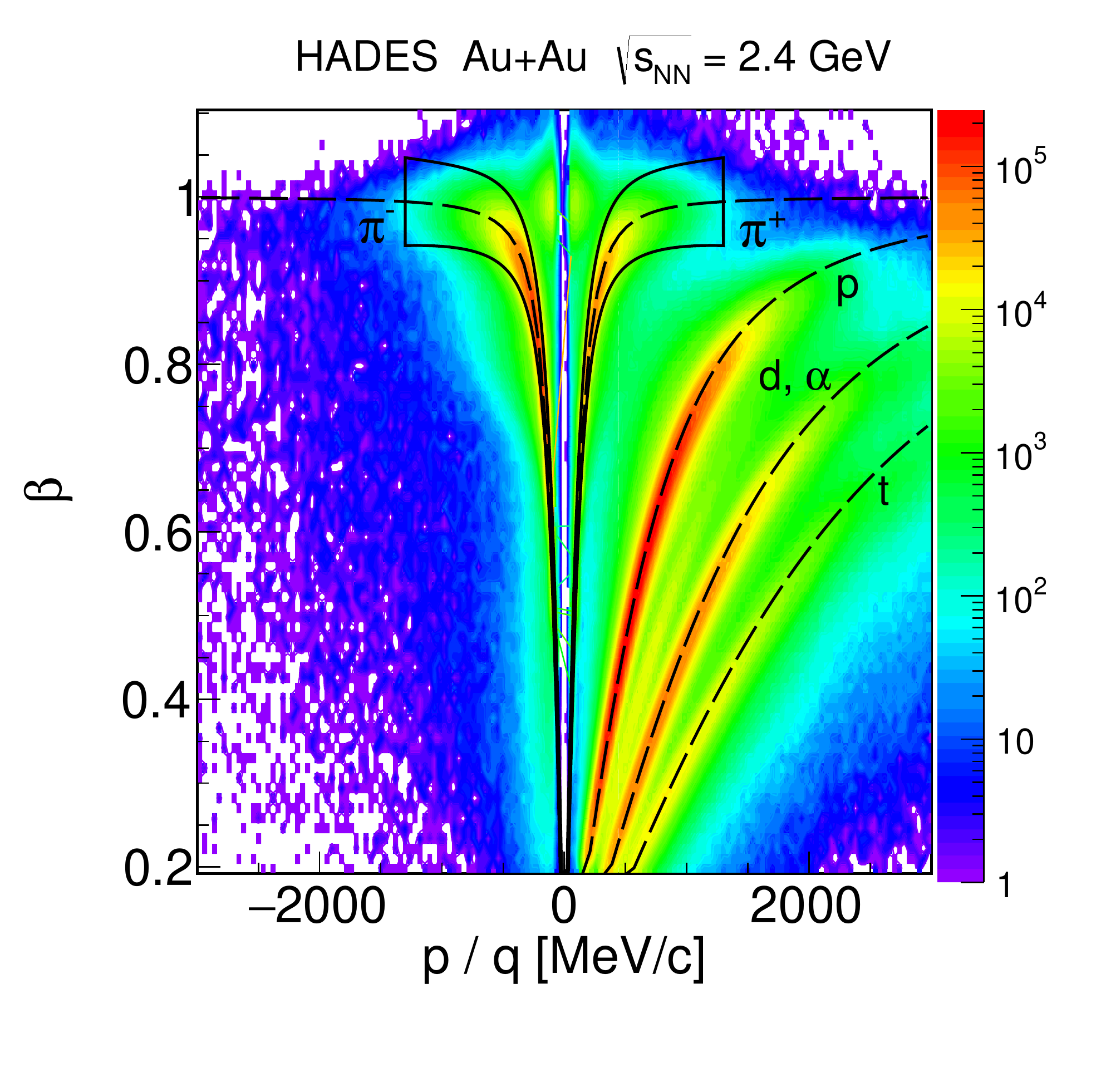}}
	\caption{Population of charged particles in the $\beta$ vs. laboratory momentum over charge ($p/q$) plane for the RPC (a) and TOF (b) detector. 
        Dashed curves correspond to the kinematic correlation for the different particle species as given by equation~\eqref{eq-beta}. 
        Solid curves show the selection of charged pions by a 2D kinematic cuts.}

	\label{fig-pid}       
\end{figure*}
\begin{figure*}
	\centering
       \subfigure[]{\includegraphics[scale=0.4]{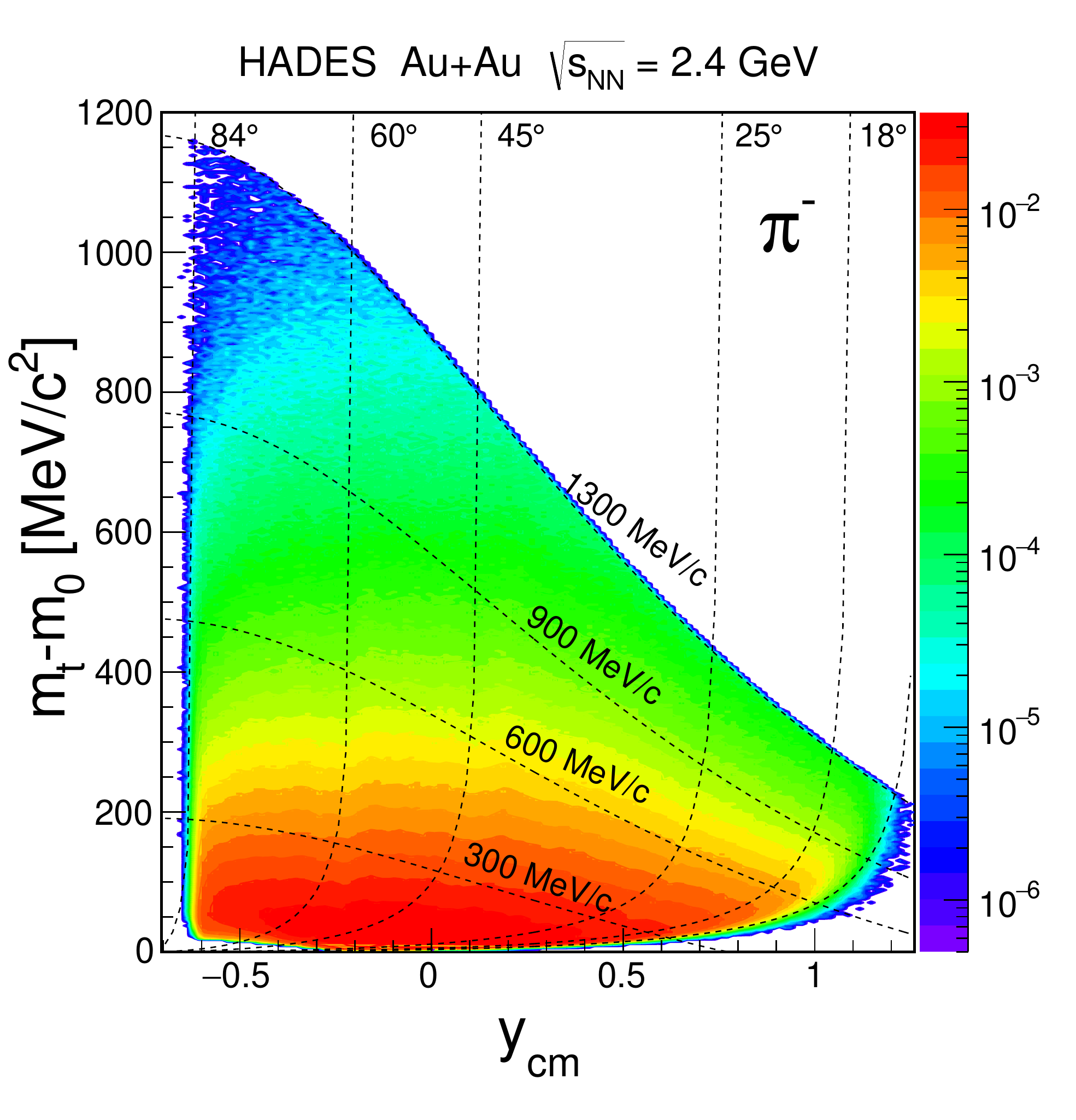}}\quad
       \subfigure[]{\includegraphics[scale=0.4]{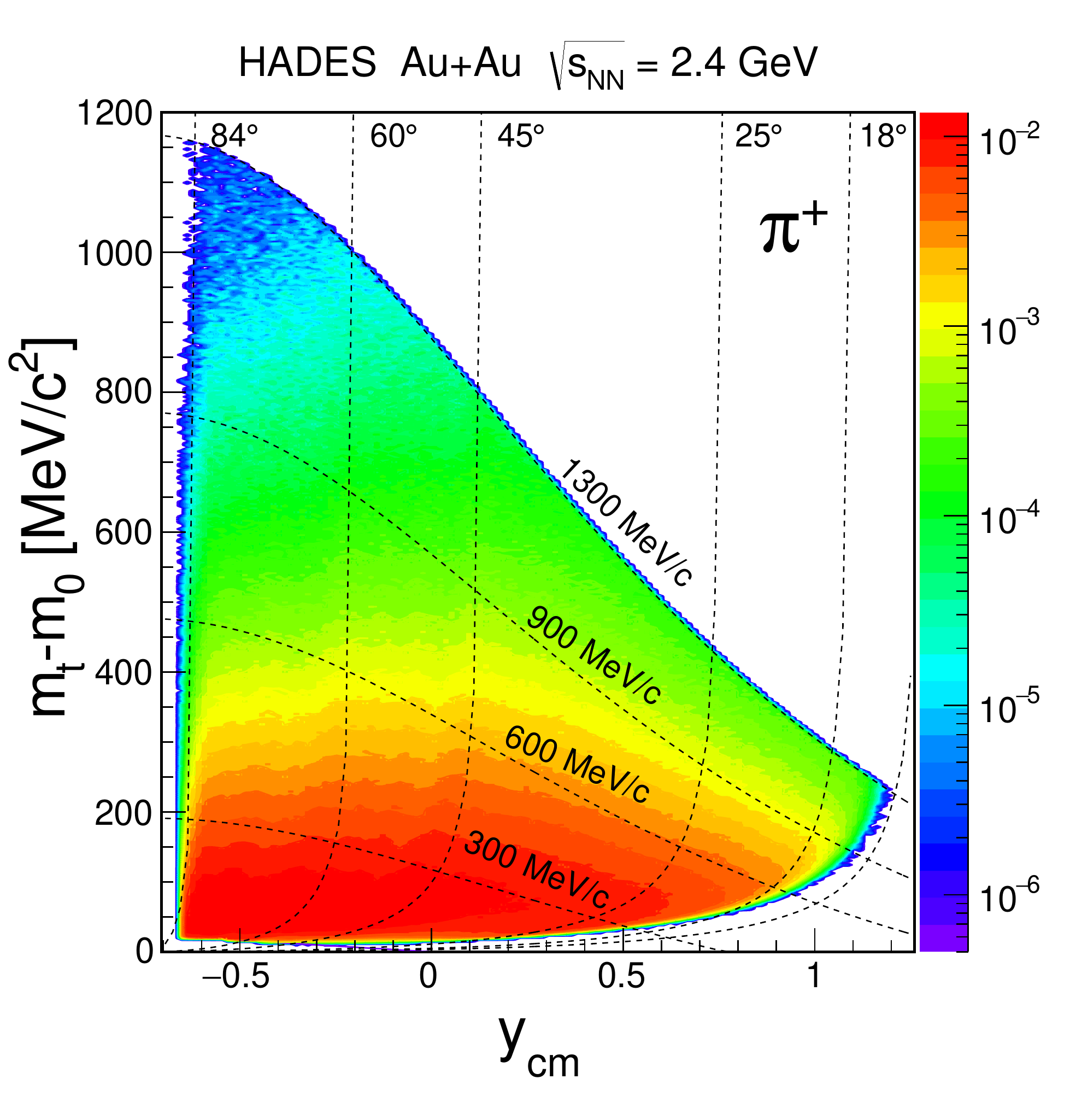}}
	\caption{Distribution of the uncorrected (raw) $\pi^-$ (a) and $\pi^+$ (b) yields in the plane spanned by rapidity ($y_{cm}$) and reduced transverse mass ($m_t -m_0$).
        The dotted curves depict laboratory polar angles and momenta. The cutoff at $p_{\pi}$ = 1.3~GeV$/c$ is caused by the PID selection (shown as solid black curve in fig.~\ref{fig-pid}).
	}
	\label{PhaseSpace}       
\end{figure*}
The experiment was performed with the High Acceptance Di-Electron Spectrometer (HADES) at the Schwerionensynchrotron SIS18 at GSI, Darmstadt. HADES, although primarily optimized to measure dielectrons 
\cite{Agakichiev:2006tg,Agakishiev:2011vf,Adamczewski-Musch:2019byl}, has also excellent hadron identification capabilities \cite{Agakishiev:2009ar,Agakishiev:2009zv,Agakishiev:2009rr,Agakishiev:2010qe,Agakishiev:2011zz,Agakishiev:2015xsa,Adamczewski-Musch:2017gjr}. 
HADES is a charged particle detector consisting of a six-coiled magnet producing a toroidal field centered around the beam axis.
Six identical detection sections are located between the coil planes and cover polar angles between 18$^\circ$ and 85$^\circ$. 
Its large azimuthal acceptance varies from 65$\%$ at low to 90$\%$ at high polar angles. 
The corresponding losses are corrected for in the analysis.  Each sector is equipped with a Ring-Imaging Cherenkov (RICH) detector for electron identification (not relevant for 
the present analysis) followed by four layers of Mini-Drift Chambers (MDCs), two in front of and two behind the magnetic field. The MDCs record space points of the trajectories of charged particles which, 
together with the known magnetic field, are used to determine the particle momentum. The momentum resolution of the charged pions was found to be $\approx 2.5~\%$ and depends only weakly on laboratory momentum. 
The arrival time of the charged particles is measured by a scintillator based Time-Of-Flight detector (TOF) covering polar angles from $44^\circ$ to  $85^\circ$ and Resistive Plate Chambers (RPC)
covering polar angles from $18^\circ$ to $45^\circ$. Their respective time resolutions are 150 and 66 ps. 
A Pre-Shower detector (behind the RPC, for electron identification) provides additional position information.
The experimental setup is described in detail in~\cite{Agakishiev:2009am}. 
\par
The beam consisted of Au$^{69+}$ ions and had an intensity of approximately $1.5\times10^6$ particles/s.  It impinged on a 15-fold segmented gold target, with an integral thickness corresponding to an interaction probability 
of 1.4 $\%$; the total length of the target assembly was 60 mm.  Several triggers were implemented: The minimum-bias trigger was defined by a valid signal in a diamond START detector in front of the target. An online trigger detected interactions and rejected peripheral 
collisions.  It was based on the summed TOF detector multiplicity signal, which selected events with more than about 20 charged particles in this detector. 
About 2.1 billion Au+Au events, corresponding to the 40\% most central collisions, were acquired this way.
The centrality of the recorded events has been mapped onto the distribution of the charged particle multiplicity $N_{ch}$ detected in the RPC and TOF detectors (for details see~\cite{Adamczewski-Musch:2017sdk}). 
Interactions of beam particles outside of the target have been rejected by requiring that the main interaction vertex is reconstructed within $\pm 32.5$ mm of the center of the target in the direction of the beam. 
The details of the procedure, which provides a measure of centrality and verifies that 
the online trigger does not bias the phase-space distributions, are described in \cite{Adamczewski-Musch:2017sdk}.
\par
Charged-hadron identification is based on the time-of-flight measured with TOF and RPC.  Particle velocity is obtained from the measured flight time and flight path.  Combining this information
with the particle momenta allows to identify charged particles (e.g. pions, kaons or protons) with high significance.  Figure~\ref{fig-pid} shows the population of all charged particles 
in the plane spanned by their $\beta$ and laboratory momenta divided by charge for the RPC (left) and TOF (right) detectors.
The different particle species are well separated and distributed along the black dashed curves which represent the function:
\begin{equation}
  \beta =  \left( \left( {\textstyle\frac{p_{lab}}{m}} \right) ^{2}+1 \right) ^{-1/2}.
\label{eq-beta}
\end{equation}
The projections of momentum slices on the $\beta$ axis are fitted with a Gaussian distribution, the mean of which is fixed to the value given by eq.~\eqref{eq-beta}. 
Their widths are free parameters which are used to select 2~$\sigma$ bands along the kinematic curves given by eq.~\eqref{eq-beta}. In order to avoid contamination of the $\pi^+$ sample by protons and of the $\pi^{\pm}$ sample 
by high momentum particles with wrong charge assignment, only pions with laboratory momenta below 1.3~GeV$/c$ are accepted for further analysis.
The losses due to the 2~$\sigma$ cut are taken care of by the efficiency and acceptance correction described below. 
The coverage in the plane spanned by rapidity (y$_{cm}$) and reduced transverse mass (m$_t$ - m$_0$)  of the measured but uncorrected yields is shown in fig.~\ref{PhaseSpace}.
\par
The measured (raw) pion spectra obtained after particle identification must be corrected for the spectrometer acceptance and losses due to the various cuts introduced during track reconstruction. 
These efficiency and acceptance corrections have been calculated from simulated Au+Au events generated by the UrQMD model~\cite{Bass:1998ca}. 
The detector response was simulated using the Geant3~\cite{Brun:1987ma} based simulation package including geometry and characteristic of all HADES detectors.
Simulations were subjected to the same reconstruction and analysis steps for all centrality classes as the experimental data. 
The fraction of lost tracks (particles) is quantified by the ratio of the number reconstructed to the number of simulated tracks. The used efficiency and acceptance correction method is described in detail 
in \cite{Agakishiev:2009zv}. The resulting correction factors were calculated in 14 rapidity ($\Delta y$ = 0.1), 60 transverse momentum ($\Delta p_t$ = 25 MeV$/c$), and 60 transverse mass ($\Delta m_t$ = 25 MeV$/c^2$) intervals (bins). 
They are typically on the order of 1.5 - 2.0 over our phase space coverage, including the correction for acceptance limitations in azimuthal angle. The bins near the acceptance limits for which the factor
was higher than 6.6 (15$\%$ efficiency) were excluded from the analysis. 
The validity of the correction procedure was checked by alternatively using a track-embedding method.  Charged pions were generated with a thermal phase space distribution and inverse slope extracted from data.
After embedding them into measured events, these events were processed by the standard reconstruction chain and the fraction of lost embedded tracks was calculated. 
It was found that the resulting correction factors differ by less than 1$\%$ from the ones obtained when using plain simulated UrQMD events. 
Differential pion yields are calculated as the product of raw yields and the correction factors in all accepted bins of $y-p_t$ and $y-m_t$ and are modeled using the following functions:
\begin{equation}
    \label{eq-boltzmann_pt}
    \frac{d^{2}N}{dp_{t}\, dy}\; =\; C_{1}(y)\, m_t\, p_t\, \exp{\textstyle \left(-\frac{m_{t}c^{2}}{ T_{1}(y) }\right)}\; +\; C_{2}(y)\, m_t\, p_t\, \exp{\textstyle \left(-\frac{m_{t}c^{2}}{ T_{2}(y) }\right)},\
\end{equation}
\begin{equation}
\label{eq-boltzmann_mt}
\frac{1}{m_{t}^{2}}\, \frac{d^{2}N}{dm_{t}\, dy}\; =\; C_{3}(y)\, \exp{\textstyle \left( -\frac{m_{t}c^{2} }{ T_{1}(y) } \right)}\; +\; C_{4}(y)\, \exp{\textstyle \left( -\frac{m_{t}c^{2} }{ T_{2}(y) } \right)}.\
\end{equation}
\begin{figure*}
	\centering
	\subfigure[]{\includegraphics[scale=0.4]{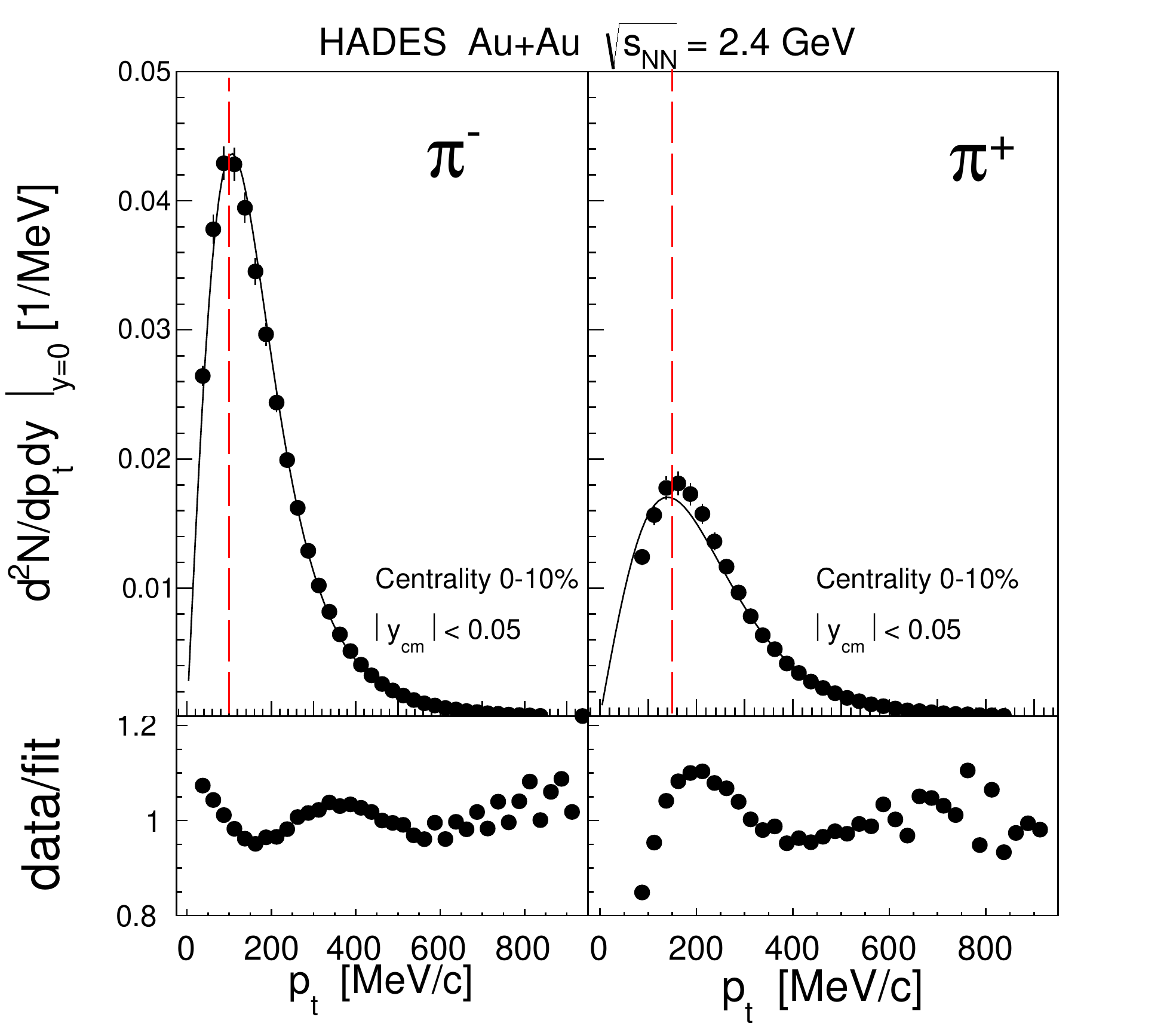}}\quad
	\subfigure[]{\includegraphics[scale=0.4]{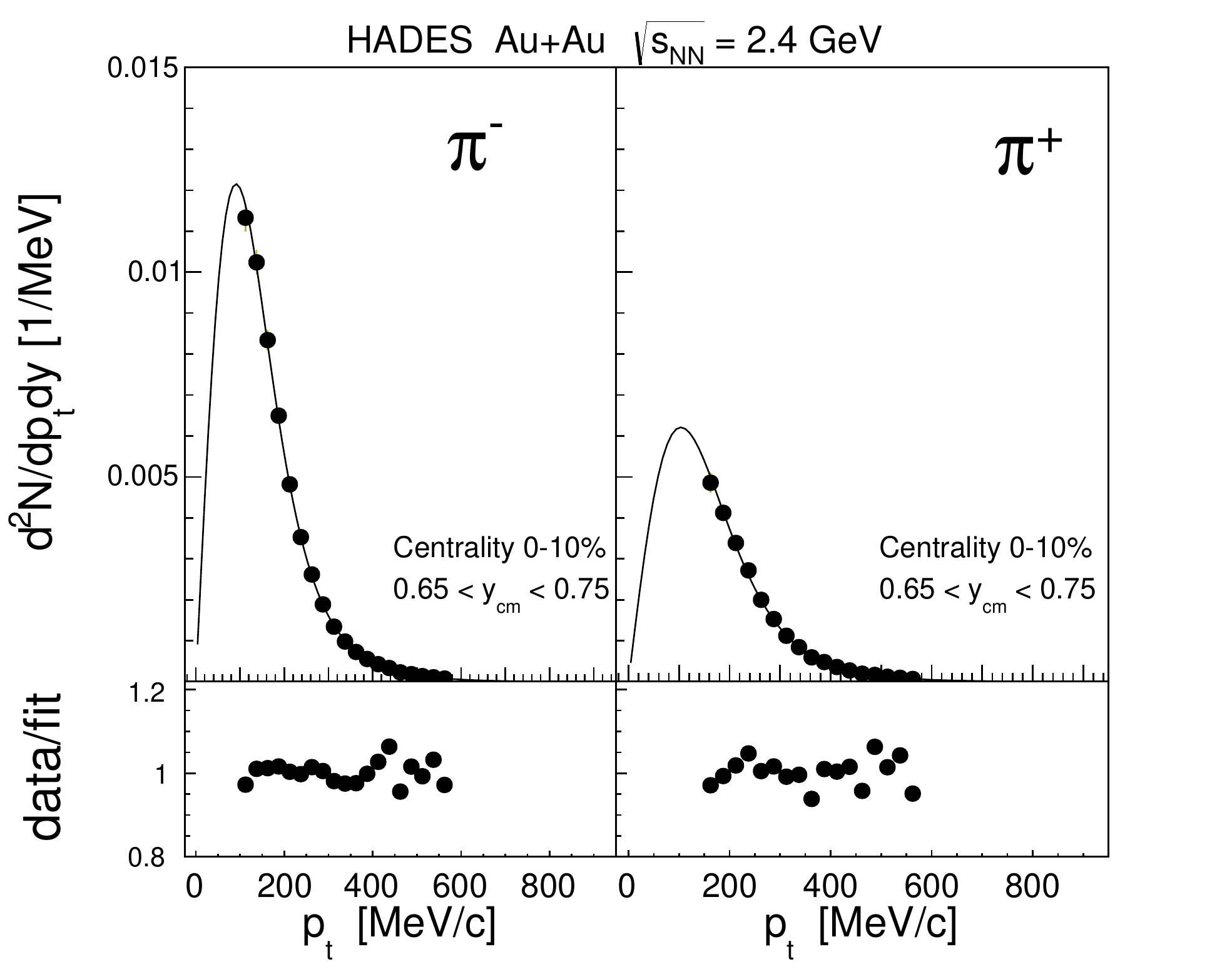}}
	\caption{
	 Mid-rapidity (a) and forward rapidity (b) transverse momentum distributions ($p_{t}$) for $\pi^+$ and $\pi^-$ mesons for the 10$\%$ most central events. The black curves represent the fit of function 
        \eqref{eq-boltzmann_pt}. The spectra are corrected for the efficiency losses and the missing acceptance in azimuth. Red dashed lines mark the
         peak position of the respective transverse momentum spectrum.  The lower part of each panel shows the ratio of data to the fitted function.}
	\label{dndpt_examples}    
\end{figure*}
\begin{figure*}
	\centering
	\subfigure[]{\includegraphics[scale=0.4]{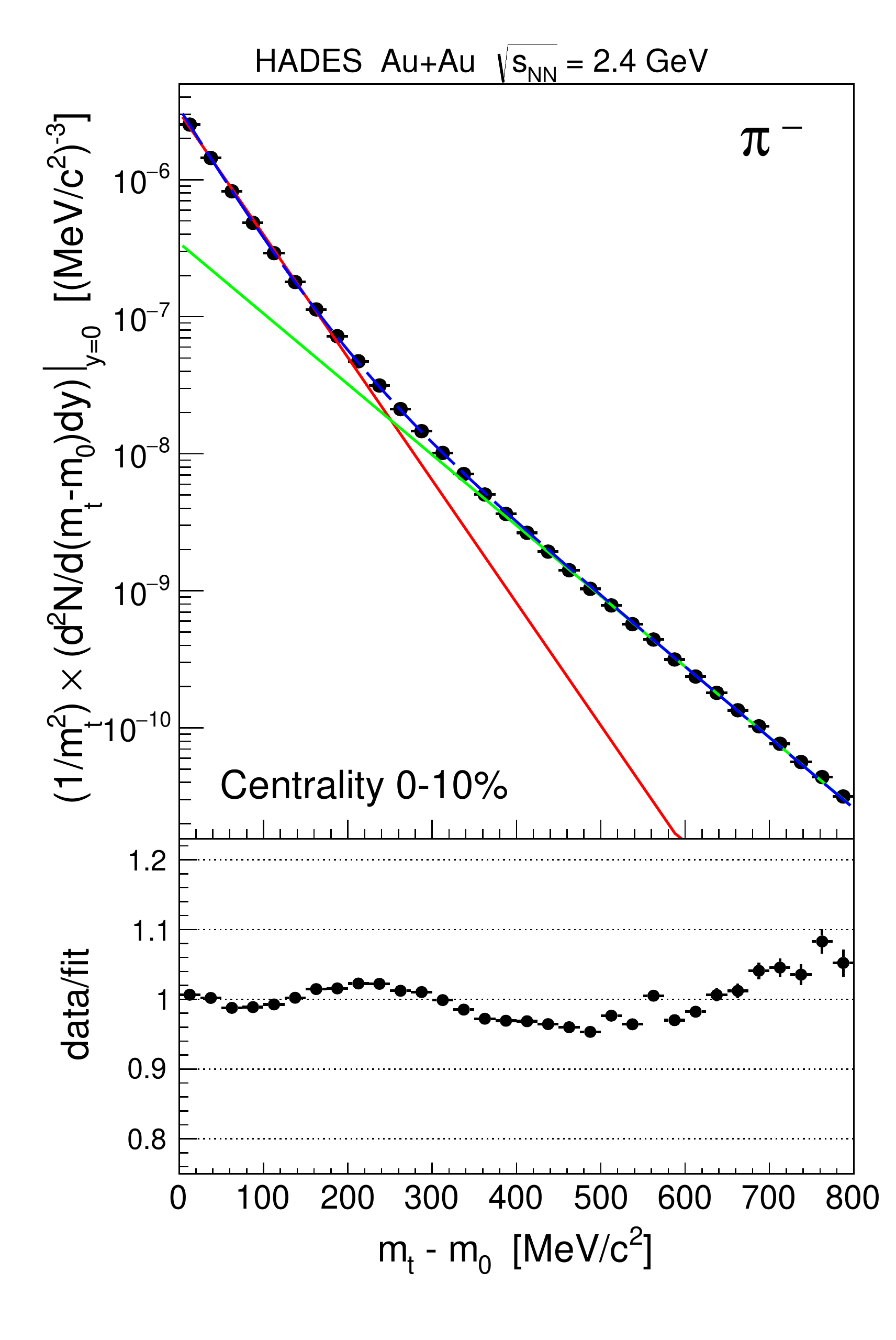}}\quad
	\subfigure[]{\includegraphics[scale=0.4]{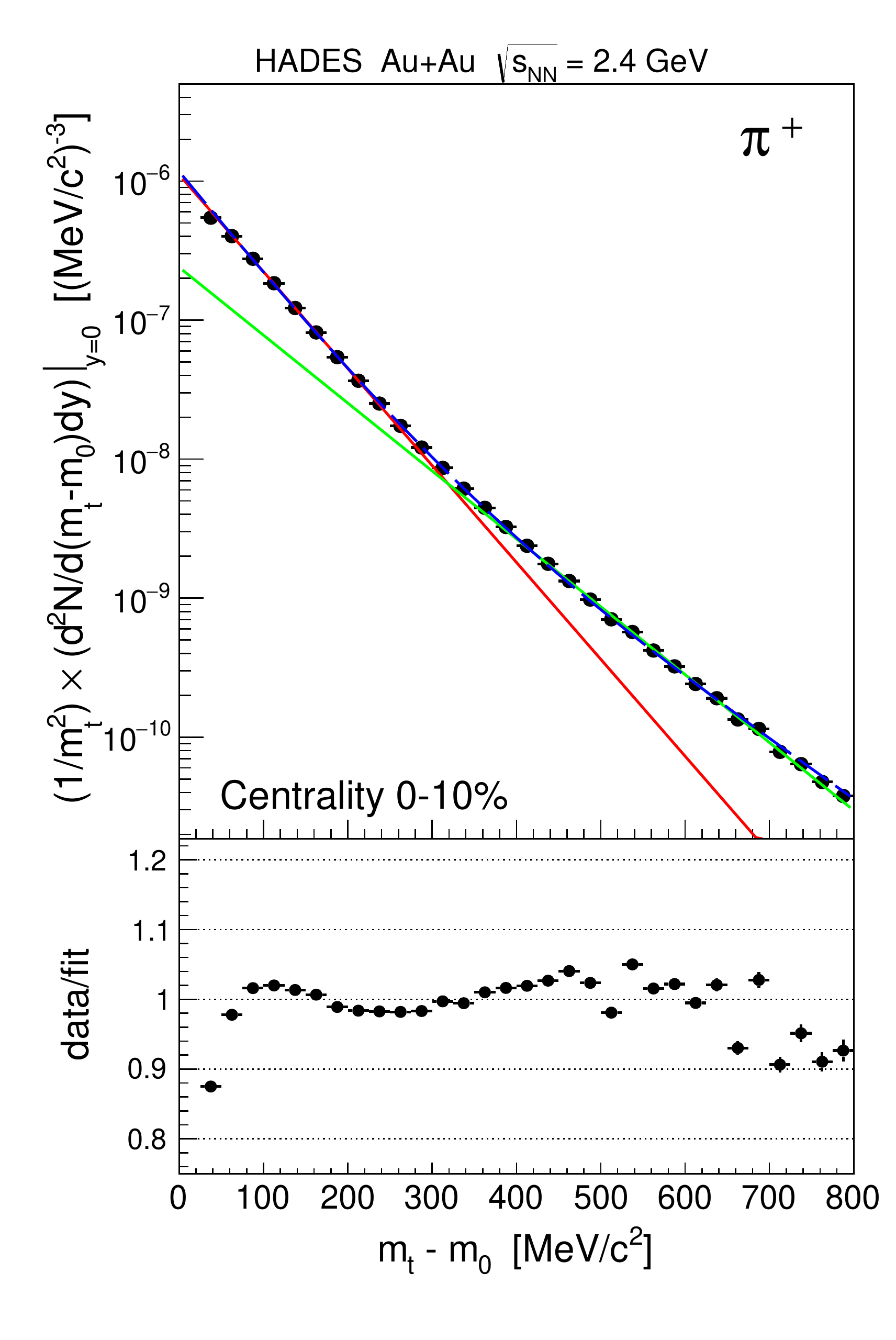}}
	\caption{
	Reduced transverse mass spectra for $\pi^-$ (a) and $\pi^+$ (b) mesons at mid-rapidity ($\mid$ y$_{cm} \mid$ $<$ 0.05) for the 10$\%$ most central events. The blue curves represent the fit using eq.~\eqref{eq-boltzmann_mt}. 
        The red and green curves represent the single Bolzmann function with the parameters $T_1$ and $T_2$ (see eq.~\eqref{eq-boltzmann_mt}). The spectra are corrected for efficiency losses and missing acceptance in azimuth.
	The lower part of each panel shows the ratio of data to the fitted double Bolzmann function.}
	\label{dndmt_examples}    
\end{figure*}
\noindent
Examples of the transverse-momentum distributions of $\pi^-$ and $\pi^+$ mesons ($dN/dp_t$) at mid-rapidity (a) and forward rapidity (b) are shown in fig.~\ref{dndpt_examples} together with 
fits of the function~\eqref{eq-boltzmann_pt}. We have chosen a superposition of two Boltzmann distributions, 
because the pion reduced transverse-mass spectra deviate from a single exponential 
as is demonstrated in figs.~\ref{dndmt_examples}~(a) and (b) which show the $m_t$ spectra at mid-rapidity. 
The parameters $T_1$ and $T_2$ account for different slopes at low and high reduced transverse masses, respectively.

The fit procedure starts with independent single Boltzmann fits in separate $m_t - m_0$ ranges 
(0-300~MeV$/c^{2}$ and 500-800~MeV$/c^{2}$). We use the resulting inverse slope parameters as starting values for $T_1$ and $T_2$ for the double Boltzmann fit in the range 0-800~MeV$/c^{2}$.
This is done in two steps. First, we require $T_2$ to be in an interval of few MeV around the value obtained from the single Boltzmann fit in order to improve the fit for $T_1$. Then we release the limits and perform the two-slope fit again, extracting the final values of $T_1$ and $T_2$, shown in table~\ref{tab-mult}.
The resulting errors are of the order of 1 MeV or smaller but depend on the chosen fit range and are correlated.  Therefore, we refrain from quoting them explicitly. 
The fit function~\eqref{eq-boltzmann_mt} is used to extrapolate the $m_t$ spectrum into the low and high regions outside of the acceptance.  The particle yield in
each rapidity interval is obtained as the sum of the measured and extrapolated  yields.  The fraction of the latter is a few percent at mid-rapidity 
and up to 30\% towards forward and backward rapidity.  The fit ranges were restricted in $p_t$ and $m_t - m_0$ to be below 800~MeV$/c^2$. 
The extrapolation into unmeasured regions is subject to systematic uncertainties.  A close look to the $p_t$ spectra of the positively (negatively) charged
pions in fig.~\ref{dndpt_examples}~(a) reveals that at low momenta the data points are systematically below (above) the fitted curves.  We attribute this deviation from
a Boltzmann shape to the Coulomb interaction of the pions with the net positive charge of the expanding fireball. 
This well-established effect \cite{Barz:1997su} causes shifts of the momentum distributions which are different for the positively and negatively charged pions. 
The former are accelerated leading to a reduced yield at low momenta and the latter are decelerated leading to an increased yield at low momenta. The comparison of
the transverse-momentum distributions of $\pi^{-}$ and $\pi^{+}$ in fig.~\ref{dndpt_examples}~(a) illustrate these momentum shifts:
the maximum of the $\pi^{-}$ ($\pi^{+}$) spectrum (represented by red dashed lines) is shifted from their average value of about 125~MeV$/c$ to 100~MeV$/c$ (150~MeV$/c$). 
Based on these results, a separate paper on the determination of the Coulomb potential is in preparation.
The deviations between the transverse-momentum (or mass) distributions and the Boltzmann fits cause a systematic underestimate (overestimate) of the extrapolation
into the low-momentum region of the $\pi^{-}$ ($\pi^{+}$) yield of 4$\%$.
The $\pi^+$ and $\pi^-$ rapidity distributions are obtained by integrating the measured transverse-mass distributions and adding the extrapolated yields (see fig.~\ref{fig-y} in \secref{sec3}). 
The missing yields in the tails of the rapidity distributions are estimated by using the $dN/dy$ shape obtained from five transport models (IQMD, PHSD, GiBUU, SMASH)
\cite{Hartnack:1997ez,Cassing:1999es,Buss:2011mx,Petersen:2018jag,Aichelin:2019tnk} (see \secref{sec4}).  The respective averaged extrapolated contributions to the yields vary
from 31\% (30\%) in central collisions to 36\% (33\%) in the peripheral ones for negatively (positively) charged pions.  The systematic uncertainty of this extrapolation
procedure is estimated by considering two extreme assumptions about the polar angle distribution of the pions (see \secref{sec3_2}): (1) The polar angle distribution of the
pions is assumed to be of the form ($1+A_2\ \cos^{2}\theta$) with $A_2$ extracted from our measurement as discussed in \secref{sec3_2}. 
This gives a lower limit which is $5\%$ smaller than the estimated yield. (2) The polar angle distribution of the pions is assumed to be of the form 
($1+A_2 \cos^{2}\theta + A_4  \cos^{4}\theta$  with $A_2$ and $A_4$ given by the shape of the polar angle distribution of $\pi^{+}$ in p+p interactions at 2~GeV 
(see fig.~19 in \cite{Fickinger:1962zz}).  This is an upper limit, because the very forward and backward preferences will be more pronounced in p+p than in $A+A$ collisions. 
This upper limit was found to be 8$\%$ and was reduced to 5$\%$ to take the higher energy of the p+p data into account. 

The statistical errors are negligible due to the large number of analyzed events.  The total systematic errors sum up to 7$\%$ based on the comparison of the corrected yield 
in the different sectors ($3\%$), as a measure of the systematic uncertainty of efficiency correction and on the errors coming from the extrapolations in rapidity ($5\%$)
and in $p_{t}$ ($m_{t}$) (4$\%$).  The different systematic uncertainties are added quadratically. 
\begin{figure*}[th]
	\begin{center}
	\subfigure[]{\includegraphics[scale=0.4]{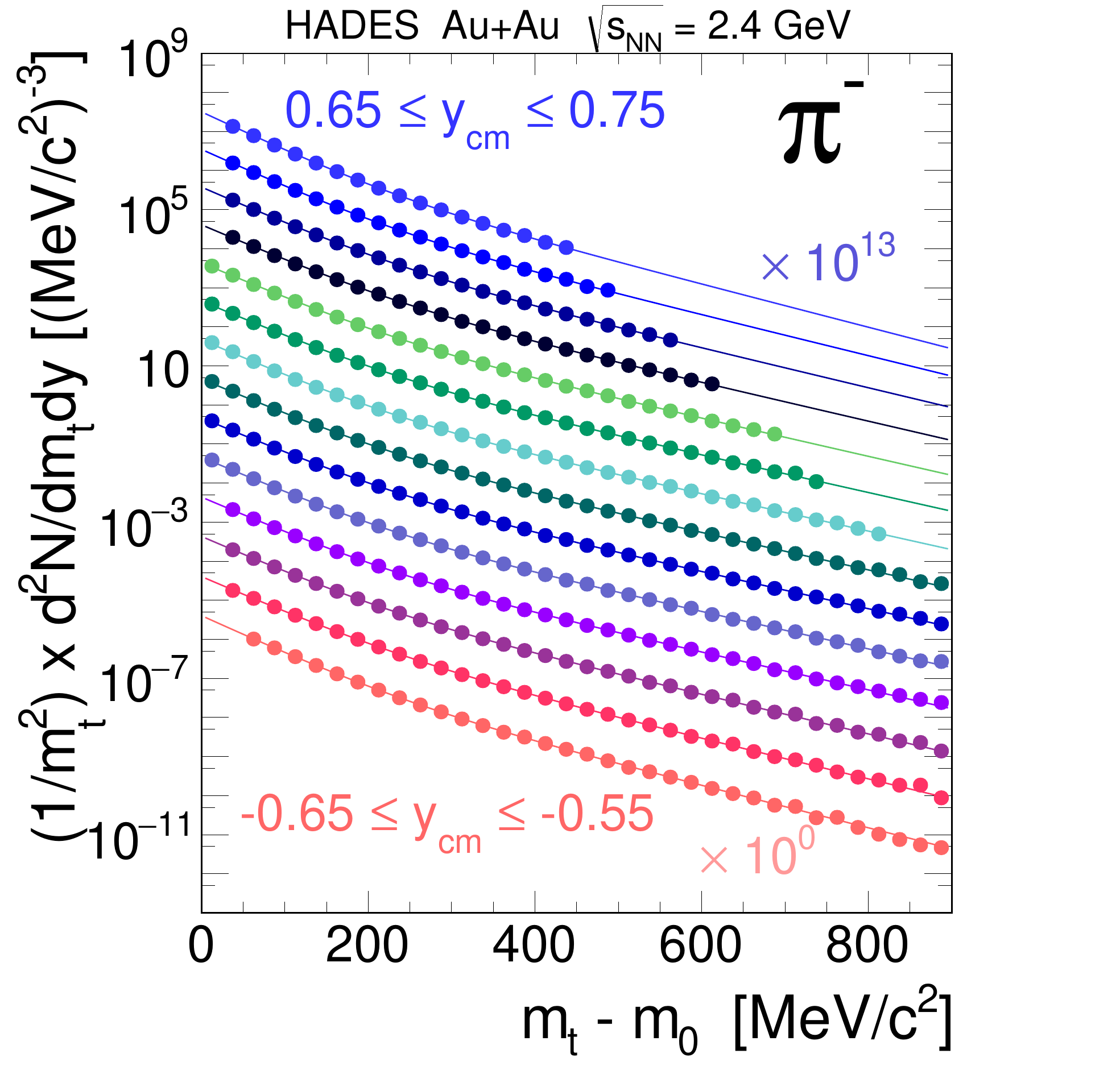}}\quad
	\subfigure[]{\includegraphics[scale=0.4]{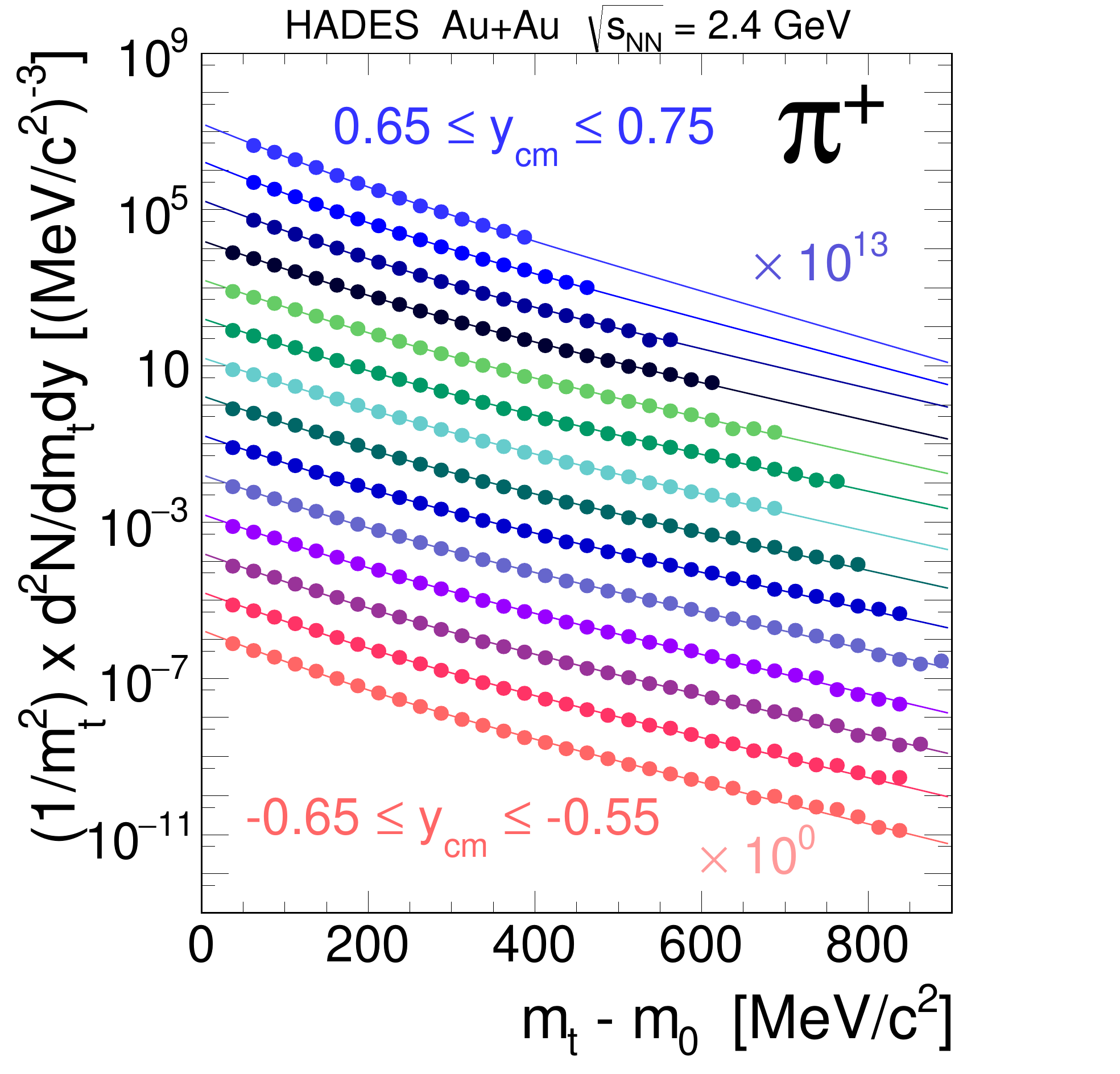}}
        \end{center}
	\caption{Reduced transverse mass distributions for negatively (a) and positively (b) charged pions in rapidity bins of $\Delta y_{cm}$ = 0.1 width
        between -0.65 and 0.75 for the 0-10$\%$ most central events. 
        The most backward rapidity data is shown unscaled while the following rapidity slices are scaled up by successive factors of 10. 
        The solid curves correspond to the two-slope Boltzmann function given by eq.~\eqref{eq-boltzmann_mt}.}	
	\label{fig-mt}      
\end{figure*}
\begin{figure*}[th]
	\centering
	\subfigure[]{\includegraphics[scale=0.4]{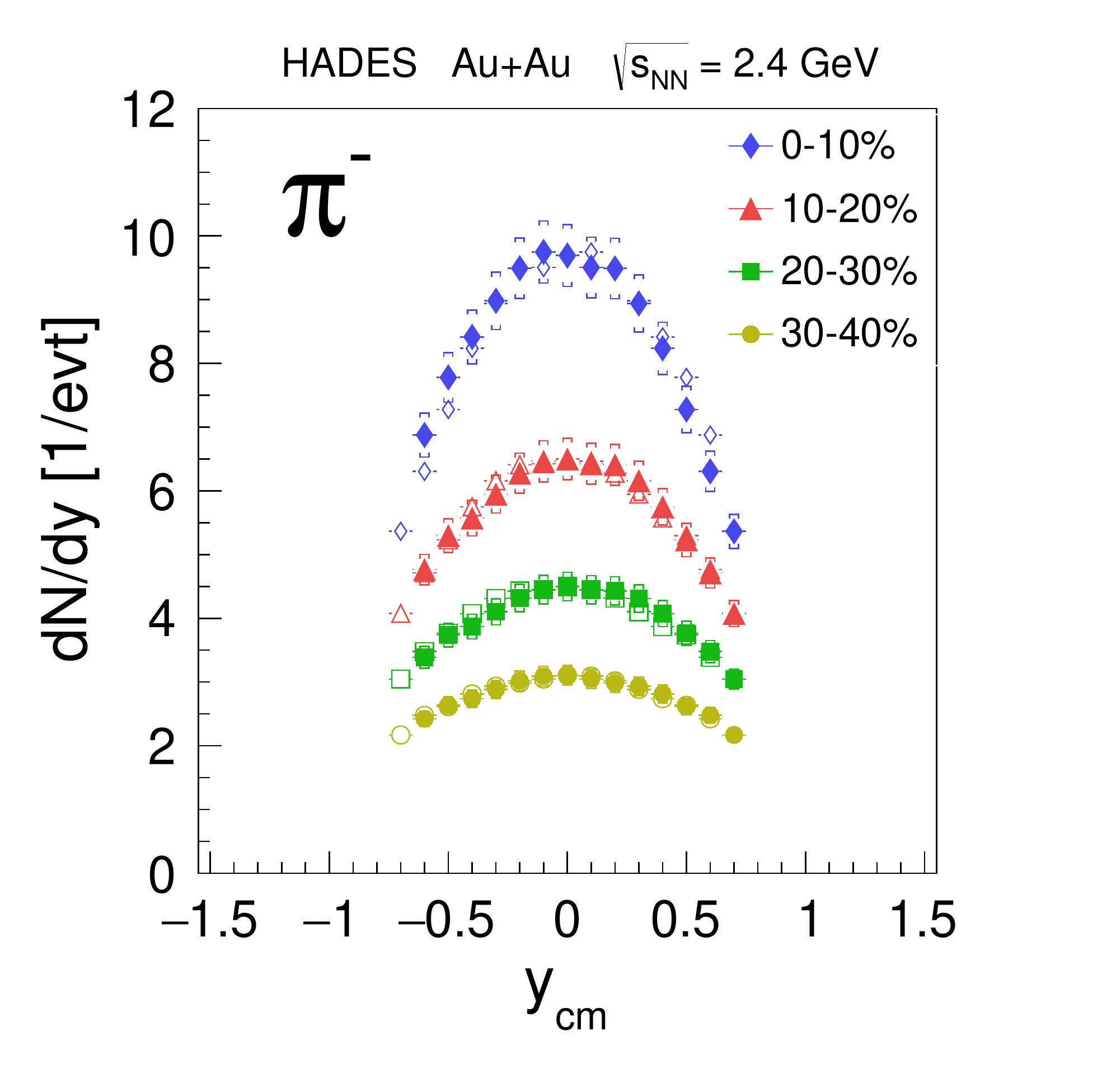}}
	\subfigure[]{\includegraphics[scale=0.4]{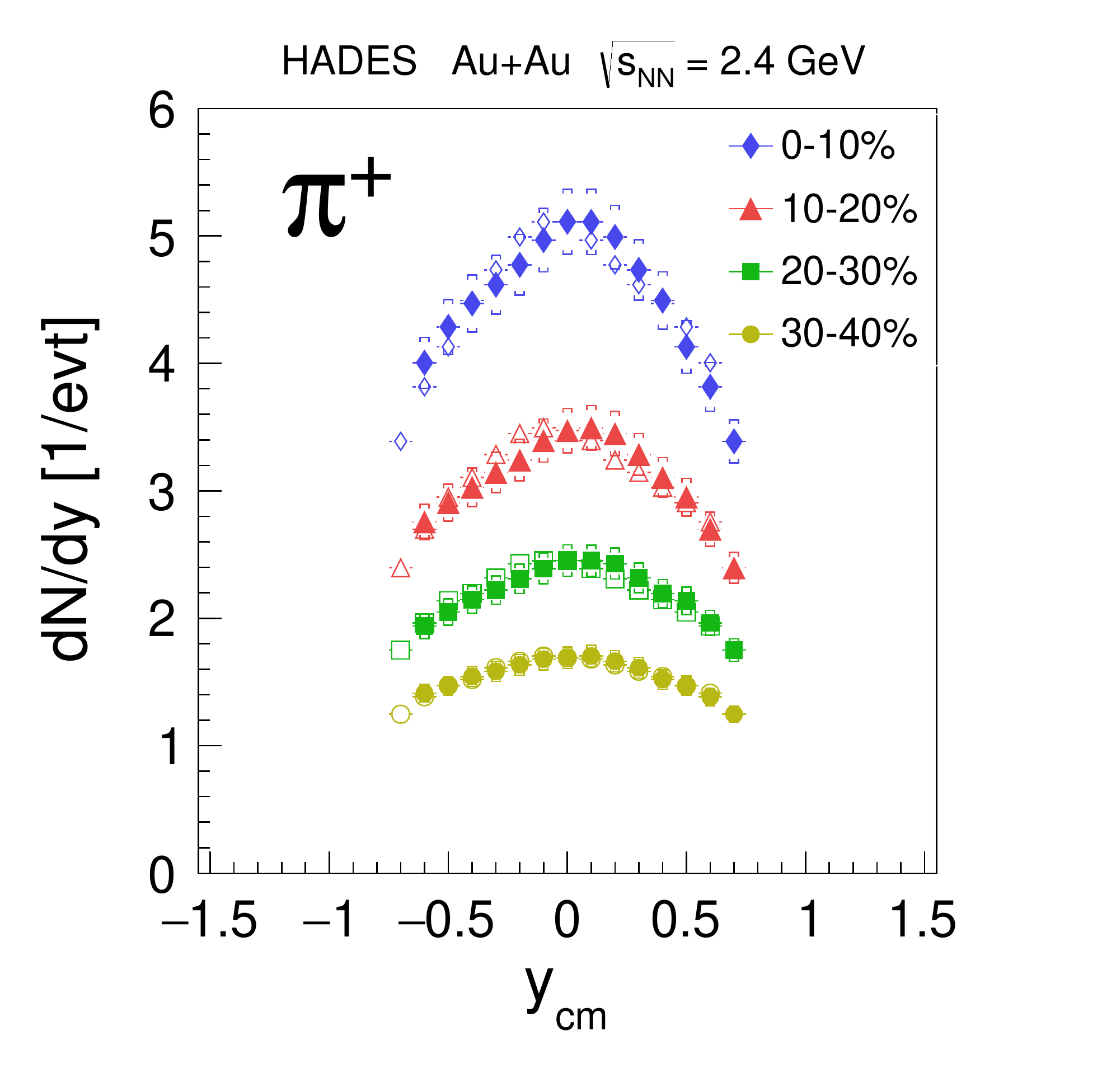}}
	\caption{Rapidity distribution of negatively (a) and positively (b) charged pions for four (10$\%$ wide) centrality classes. 
		Full points are the measured data and open points are the data reflected at y$_{cm}$ = 0. The error bars indicate the systematic uncertainties. The statistical errors are negligible.}
	\label{fig-y}       
\end{figure*}

\section{Results}
\label{sec3}
\begin{table}
	\centering
	\begin{tabular}{ cccccc }
		\hline
		$\pi^{-}$ &  yield            & yield 4$\pi$      & $T_{1}$  & $T_{2}$ \\
                          &    [1/evt]   & [1/evt]    &   [MeV]  &  [MeV]                \\ \hline
		0-40$\%$  &    7.3 $\pm$ 0.4  & 11.1 $\pm$ 0.6 $\pm$ 0.6  & 44                & 87               \\ \hline
		0-10$\%$  &   11.6 $\pm$ 0.6  & 17.1 $\pm$ 0.9 $\pm$ 0.9  & 46                & 91               \\
		10-20$\%$ &    8.0 $\pm$ 0.4  & 12.1 $\pm$ 0.6 $\pm$ 0.6  & 43                & 85               \\
		20-30$\%$ &    5.6 $\pm$ 0.3  &  8.7 $\pm$ 0.4 $\pm$ 0.4  & 42                & 82               \\
		30-40$\%$ &    3.9 $\pm$ 0.2  &  6.3 $\pm$ 0.3 $\pm$ 0.3  & 41                & 79               \\ \hline\hline
		
		$\pi^{+}$ & yield    & yield 4$\pi$      & $T_{1}$  & $T_{2}$   \\ 
                          &    [1/evt]   & [1/evt]    &   [MeV]  &  [MeV]                \\ \hline
		0-40$\%$  &   3.9 $\pm$ 0.2   &  6.0 $\pm$ 0.3 $\pm$ 0.3   & 52                &  88               \\ \hline
		0-10$\%$  &   6.2 $\pm$ 0.3   &  9.3 $\pm$ 0.5 $\pm$ 0.5   & 54                &  92               \\
		10-20$\%$ &   4.3 $\pm$ 0.2   &  6.6 $\pm$ 0.3 $\pm$ 0.3   & 51                &  89               \\
		20-30$\%$ &   3.0 $\pm$ 0.2   &  4.7 $\pm$ 0.2 $\pm$ 0.2   & 49                &  86               \\
		30-40$\%$ &   2.1 $\pm$ 0.1   &  3.4 $\pm$ 0.2 $\pm$ 0.2   & 47                &  83               \\
		
		\hline
	\end{tabular}
\vspace{0.5cm}
        \caption[]{Measured ("yield") and extrapolated ("yield 4$\pi$") particle multiplicities for four (10$\%$ wide) centrality classes and for 0-40$\%$ range. 
		The statistical errors are negligible. Shown are systematic uncertainties due to the correction procedure and the extrapolation in $m_t$ (added quadratically, first error) and extrapolation in rapidity 
                (second error). In addition, the two inverse slope parameters $T_{1}$ and $T_{2}$ for mid-rapidity are listed. Here errors are omitted, because both parameters depend on the chosen fit range and are correlated.}
        		\label{tab-mult}

\end{table}
The complete set of the corrected charged-pion measurements is given in fig.~\ref{fig-mt} together with the fits corresponding to eq.~\eqref{eq-boltzmann_mt}.
The resulting slope parameters $T_1$ and $T_2$ of the transverse-mass spectra at mid-rapidity are listed in table~\ref{tab-mult}. $T_1$ describes the slope of the
low $m_t$ part of the spectrum which contains the bulk of the particles and is usually attributed to pions originating from $\Delta$ decays.  $T_2$ stands for the
slope at higher $m_t$ which is often interpreted as a thermal component \cite{Hong:1997ka}, but can be also attributed to pions from decays of various broad higher-lying resonances. 

The rapidity distributions of $m_t$ extrapolated and integrated yields for both charges are presented in fig.~\ref{fig-y} for the centrality classes 0-10\%, 10-20\%,
20-30\%, and 30-40\%.  The $4\pi$ yields and their errors listed in table~\ref{tab-mult} refer to the means and the scatter of the values obtained from the extrapolations
in rapidity described in the previous section. 

\begin{figure*}
	\centering
	\subfigure[]{\includegraphics[scale=0.4]{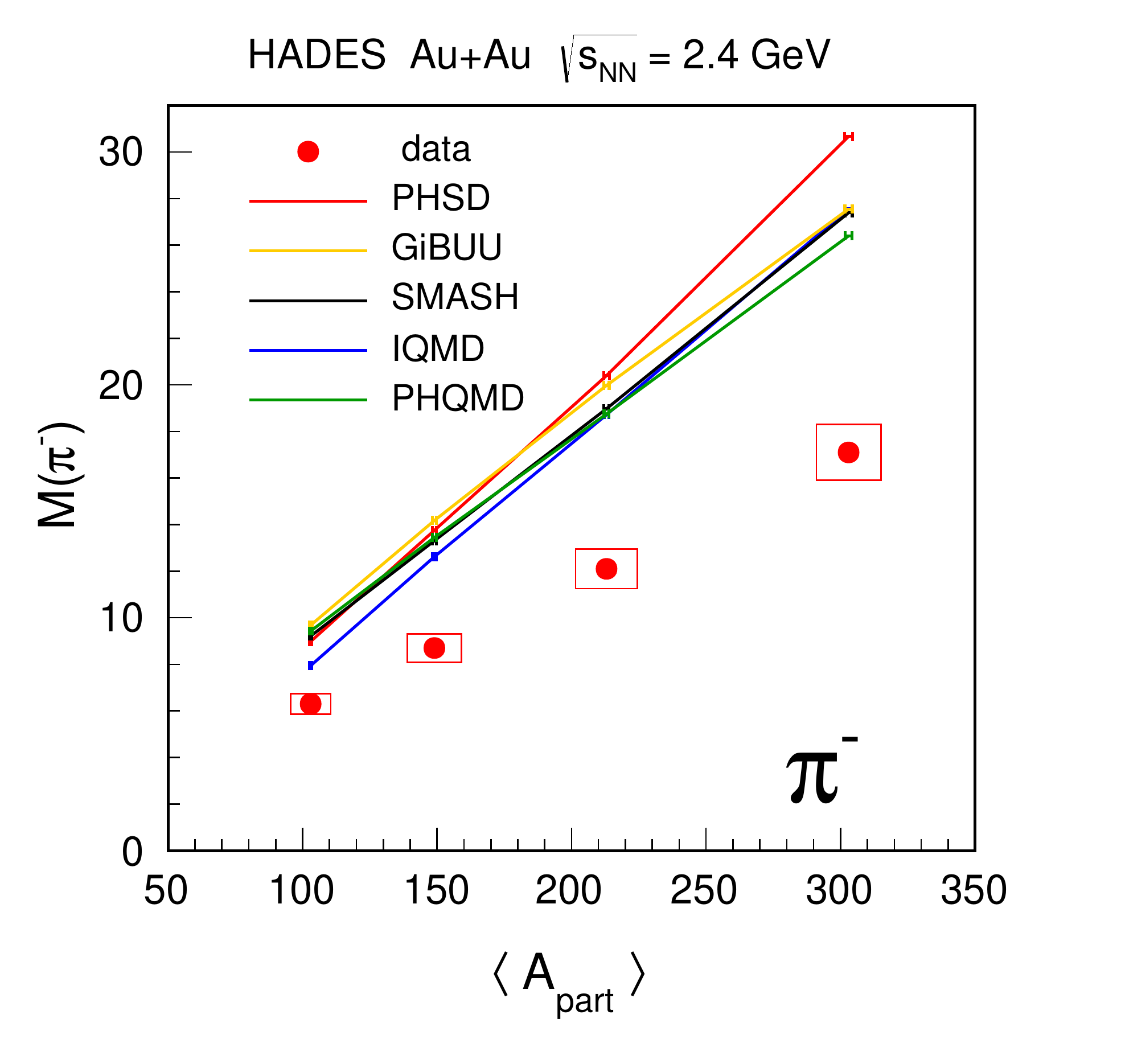}}\quad
	\subfigure[]{\includegraphics[scale=0.4]{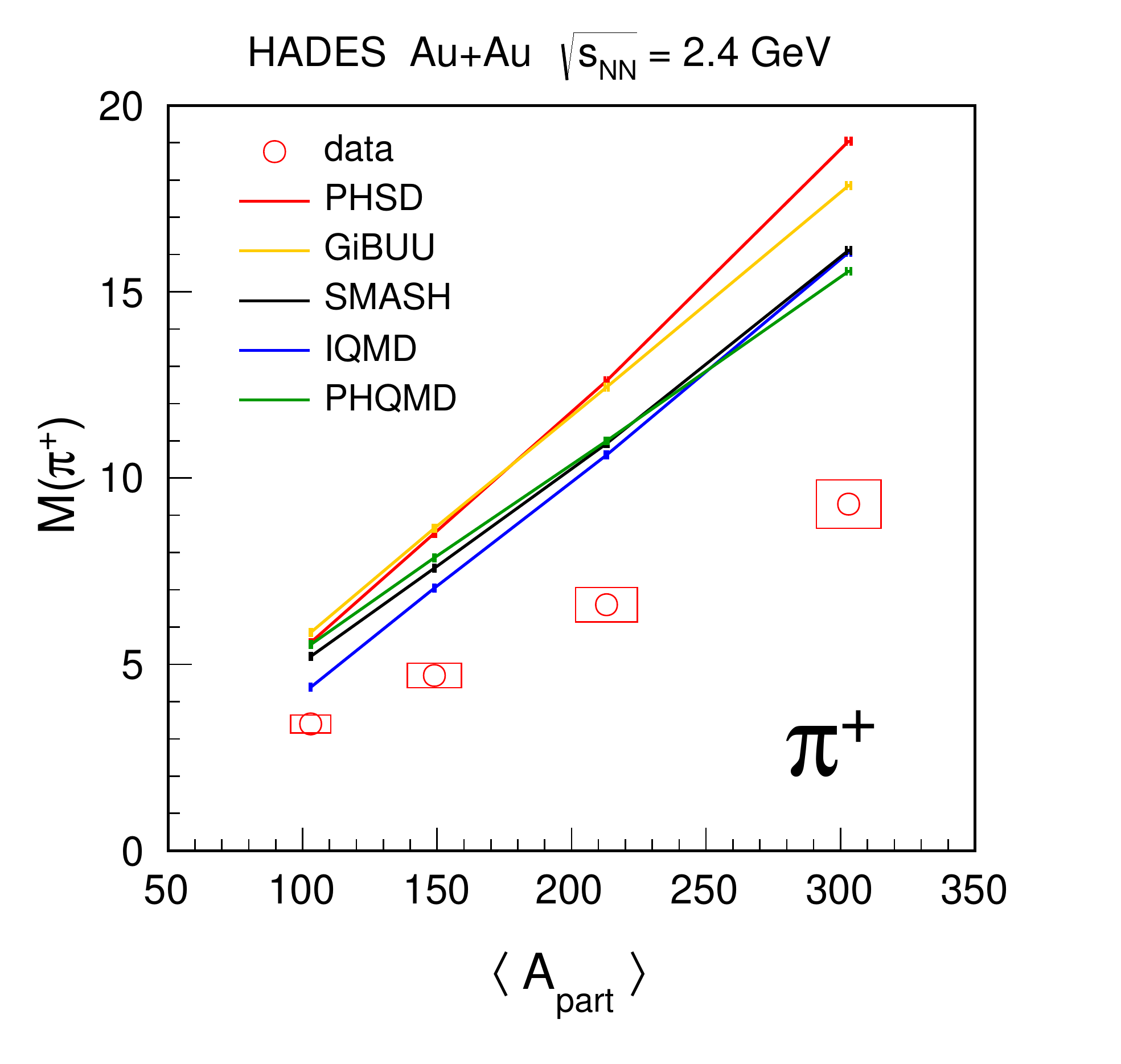}}
	\caption{Multiplicities of $\pi^{-}$ (a) and $\pi^{+}$ (b) as a function of the  mean number of participants $\langle A_{part}\rangle$. The vertical size of the open boxes stands 
        for the systematic uncertainties due to the correction factors and extrapolations. The horizontal size of the open boxes indicates the error on the mean number of participants (for details see~\cite{Adamczewski-Musch:2017sdk}).
	Colored solid lines represent the results of various model calculations for $\pi^-$ ($\pi^+$) (see \secref{sec4}).
	}
	\label{fig-MultApart}     
\end{figure*}

Figure~\ref{fig-MultApart} shows the centrality dependence of pion yields in Au+Au collisions with the centrality parameterized by the mean number of participants $\langle A_{part}\rangle$. 
The pion multiplicity per participating nucleon as a function of decreasing centrality and system size increases in our data from 0.13 in the most central Au+Au collisions (0-10$\%$)
to 0.14 in the 30-40$\%$ interval, see also fig. \ref{fig-world-Apart} and fig \ref{fig:ApartSystems-Ebeam}. 
Note that the statistical errors are negligible and most of the systematic uncertainties partially cancel when comparing the multiplicities in different centrality
classes relative to each other.  The scaling of the yields with the mean number of participants $\langle A_{part}\rangle$ is quantified by the scaling parameter $\alpha$, 
where the multiplicity $M(\pi^{\pm})$ is $\propto \langle {A_{part}} \rangle ^\alpha$.  We find for $\pi^-$ a value of $\alpha= 0.93\pm0.01$ and for $\pi^+$ a value of $\alpha= 0.94\pm0.01$. Hence, we observe for both pion species a significantly weaker than linear scaling with the mean number of participants $\langle A_{part}\rangle$.
Hence, we observe for both pion species a significantly weaker than linear scaling with the mean number of participants $\langle A_{part}\rangle$ and the decrease
with increasing centrality observed in fig.~\ref{fig-world-Apart} is indeed significant.

The multiplicity of $\pi^{-}$ mesons is larger than the one of the $\pi^{+}$ due to the neutron over proton excess in the Au nucleus. 
Using the parameterizations from~\cite{VerWest:1981dt} for the energy dependence of the pion production cross sections in the different isospin channels of nucleon-nucleon
interactions, the $\pi^{-} / \pi^{+}$ ratio can be calculated for our beam energy (1.23~A~GeV).  
The respective value of $1.84$  agrees with the experimental finding of $1.83 \pm 0.17.$
\subsection{Beam-energy and system-size dependence}
\label{sec3_1}
\begin{figure}[]
	\centering
	\includegraphics[width=1.\linewidth]{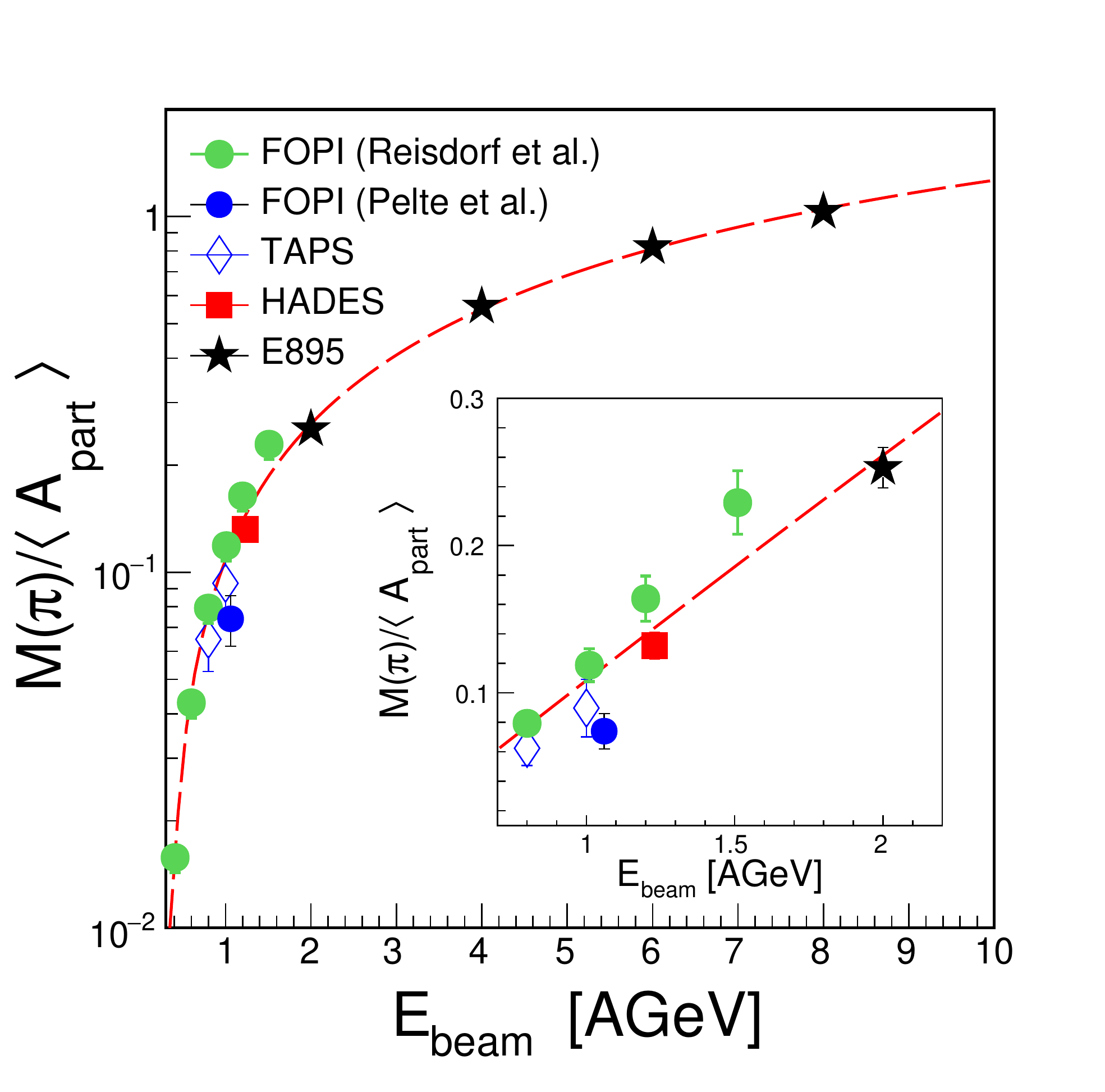}
	\caption{Pion multiplicity $M(\pi)$ per mean number of participating nucleon $\langle A_{part}\rangle$ as a function of the kinetic beam energy $E_{beam}$.	
                 The dashed curve is a fit to the data points except for the one labeled "FOPI (Pelte et al.)", as suggested in \cite{Reisdorf:2006ie}. 
                 The inset magnifies the energy region around the HADES point. }
    \label{fig-world-Ebeam}  
\end{figure}

\begin{figure}[]
	\centering
	\includegraphics[width=1.\linewidth]{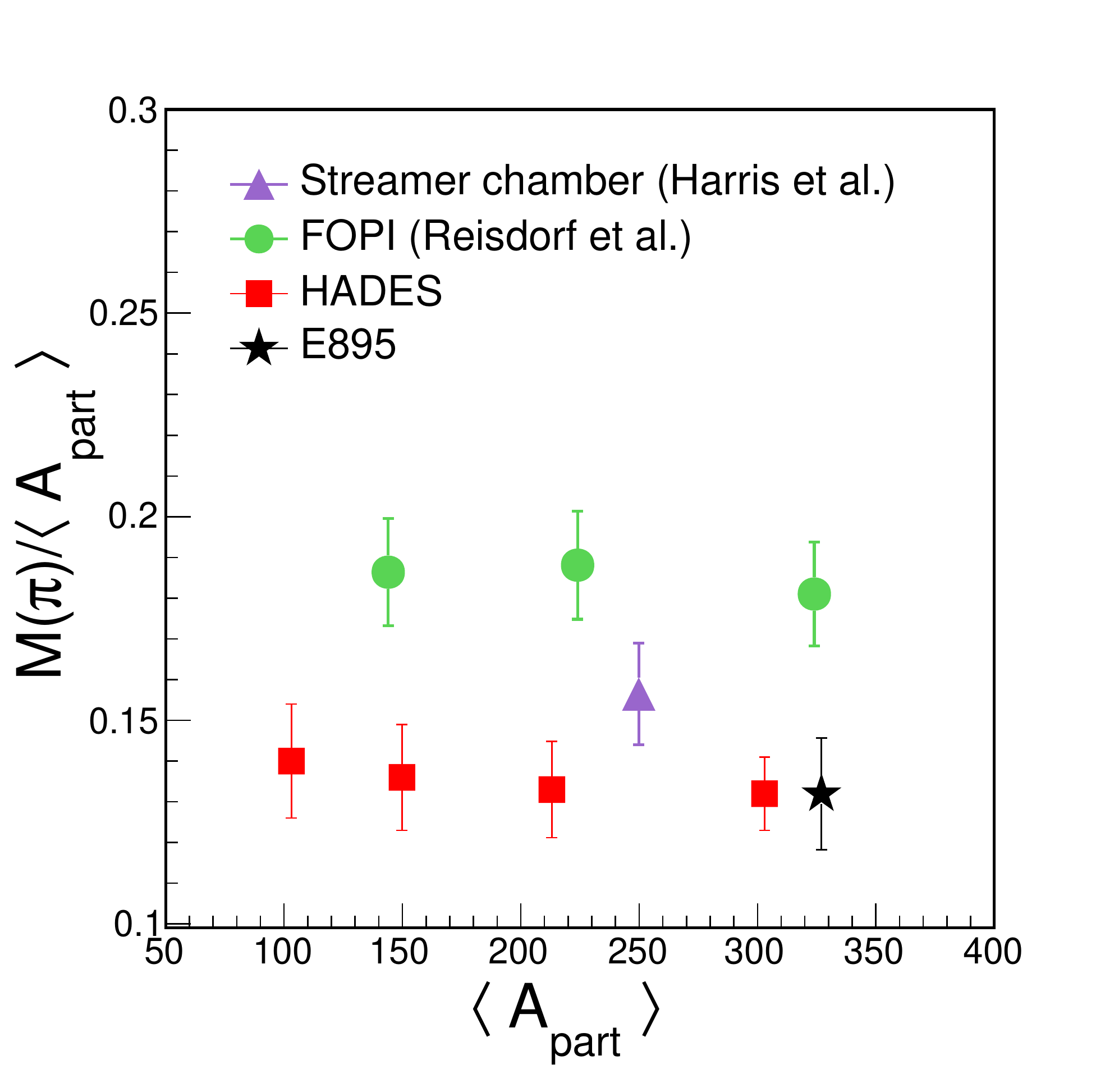}
	\caption{Comparison of the centrality dependence of $M(\pi)/\langle A_{part}\rangle$ in Au+Au collisions 
                 to earlier measurements at similar energies. The results from FOPI, E895, and from the BEVALAC Streamer Chamber group (the latter for $La+La$ collisions) 
                 have been scaled to 1.23~A~GeV; note the suppressed zero on the ordinate.
    }
    \label{fig-world-Apart}  
\end{figure}

The excitation function of pion multiplicities in Au+Au collisions from threshold up to beam energies of 10~A~GeV is displayed in fig.~\ref{fig-world-Ebeam} and fig.~\ref{fig-world-Apart}. In the past, in this energy range pion data have been collected by 
the TAPS~\cite{Wolf:1998vn,Averbeck:2000sn}, the KaoS~\cite{Wagner:1997sa}, the FOPI~\cite{Reisdorf:2006ie,Pelte:1997rg} and the E895~\cite{Klay:2003zf} experiments. 
The world data, together with the presented result, are plotted in terms of the normalized total pion multiplicity $M(\pi)/\langle A_{part}\rangle$ as a function of the beam
kinetic energy $E_{beam}$ where $M(\pi) = M(\pi^{+}) + M(\pi^{-}) + M(\pi^{0})$ and $M(\pi^{0})$, when not available, is approximated as $M(\pi^{0}) = 0.5 \times [M(\pi^{+}) + M(\pi^{-})]$. 
The number of participants $ A_{part}$ is not a direct observable and the methods used for its estimation vary between different experiments, which puts limits on the accuracy
of such a comparison (Note that all published TAPS data were extrapolated to 4$\pi$ assuming isotropic emission from a source at mid-rapidity).  Our results for $M(\pi)$ are
listed in table~\ref{tab-Apart-b} (together with values of $\langle A_{part}\rangle$ taken from~\cite{Adamczewski-Musch:2017sdk}) and fit well into the overall systematics of
the world data.  
\begin{table}
	\centering
	\begin{tabular}{cccc}
		\hline
		          & $M(\pi)$                    & $\langle A_{part}\rangle$ & b $[fm]$   \\ \hline
		0-40$\%$  &  25.6 $\pm$  1.8  & 193 $\pm$ 13              &  0.0 $-$ 9.3       \\ \hline
		0-10$\%$  &  40.0 $\pm$  2.8  & 303 $\pm$ 12              &  0.0 $-$ 4.7       \\
		10-20$\%$ &  28.4 $\pm$  2.0  & 213 $\pm$ 12              &  4.7 $-$ 6.6       \\
		20-30$\%$ &  20.2 $\pm$  1.4  & 149 $\pm$ 10              &  6.6 $-$ 8.1       \\
		30-40$\%$ &  14.4 $\pm$  1.0  & 103 $\pm$  8              &  8.1 $-$ 9.3       \\ 
				
		\hline
	\end{tabular}
\vspace{0.5cm}
	\caption[]{Total pion multiplicities for four (10$\%$ wide) centrality classes and for the full 0-40$\%$ range.  Listed are also the mean number of participants $\langle A_{part}\rangle$ (taken from~\cite{Adamczewski-Musch:2017sdk})
        and impact parameter ranges used for centrality selection in the models.
	}
	\label{tab-Apart-b}
\end{table}
We model the energy dependence of the pion multiplicity with a simple second order polynomial ($a_{0} + a_{1}E_{beam} +a_{2}E_{beam}^{2}$)
(the dashed line in fig.~\ref{fig-world-Ebeam}). The resulting parameters are $a_{0}=-5.36 \times 10^{-2}$, $a_{1}=1.72 \times 10^{-1}$~A~GeV$^{-1}$, $a_{2}=-3.08 \times 10^{-3}$~A~GeV$^{-2}$. It turns out that the early result from the FOPI experiment 
at 1~A~GeV \cite{Pelte:1997rg} is significantly below the value suggested by the world data. This has already been discussed by the FOPI collaboration in~\cite{Reisdorf:2006ie} and attributed to detector effects. 
Therefore we exclude this measurement in the fit. Furthermore, the FOPI data points at 1.2 and 1.5~A~GeV are 25\% (2.5$~\sigma$) above our data and the trend of the world data. 
\begin{figure}
	\centering
	\includegraphics[width=1.\linewidth]{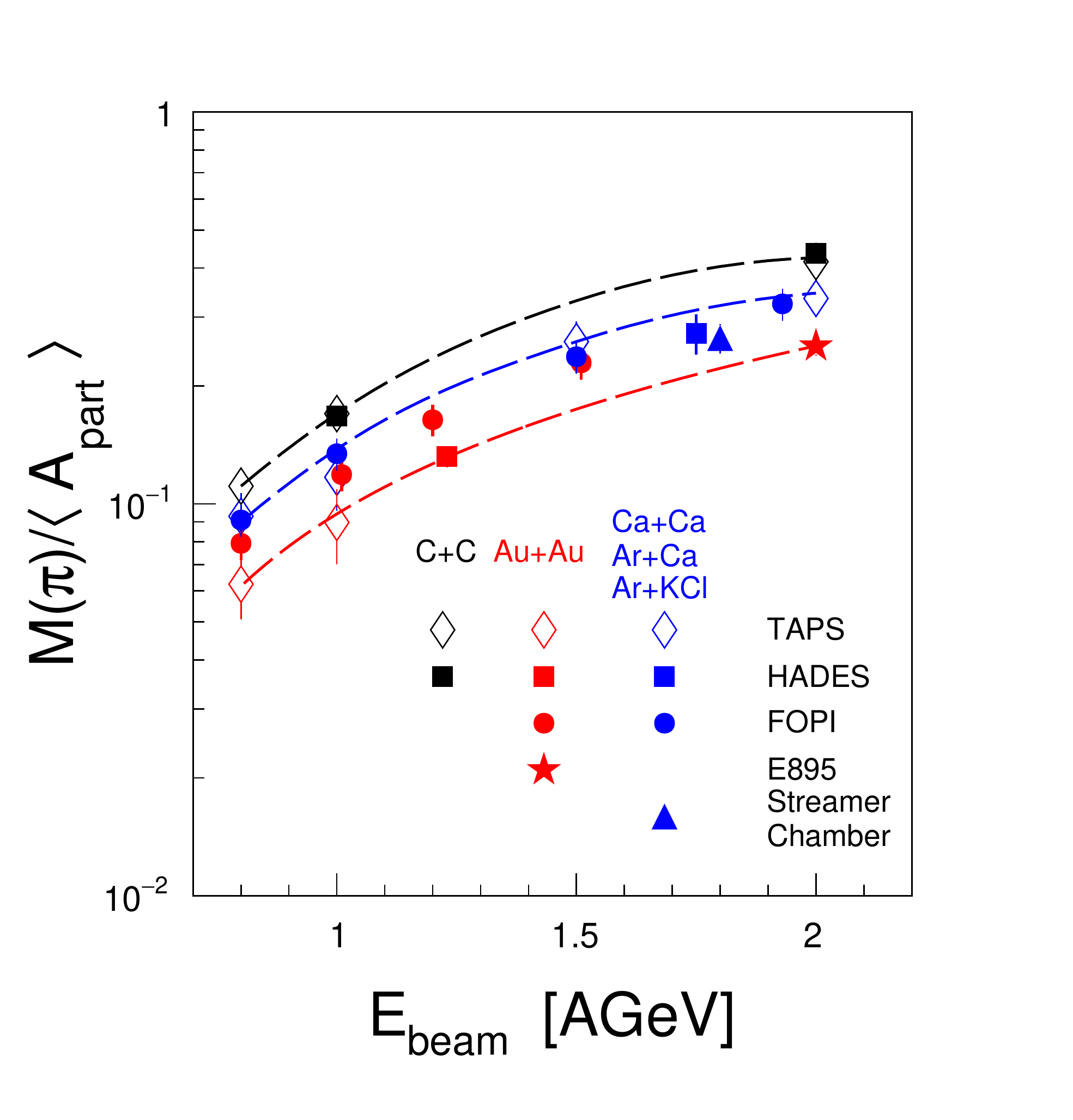}
    \caption{Pion multiplicity per participating nucleon as a function of beam energy for three different systems:
    C+C (black) \cite{Agakishiev:2007ts,Agakichiev:2006tg,Averbeck:2000sn}, Ar+KCl (blue) \cite{Averbeck:2000sn,Averbeck:1997ma,Holzmann:1997mu,Reisdorf:2006ie,Sandoval:1980bm} and Au+Au (red) \cite{Wolf:1998vn,Averbeck:2000sn,Reisdorf:2006ie,Klay:2003zf}. 
    The curves are polynomial fits to these data used to interpolate the multiplicities as a function of bombarding energy
    for corresponding systems.}
    \label{fig:ApartSystems-Ebeam}       
\end{figure}
\begin{figure}
	\centering
	\includegraphics[width=1.\linewidth]{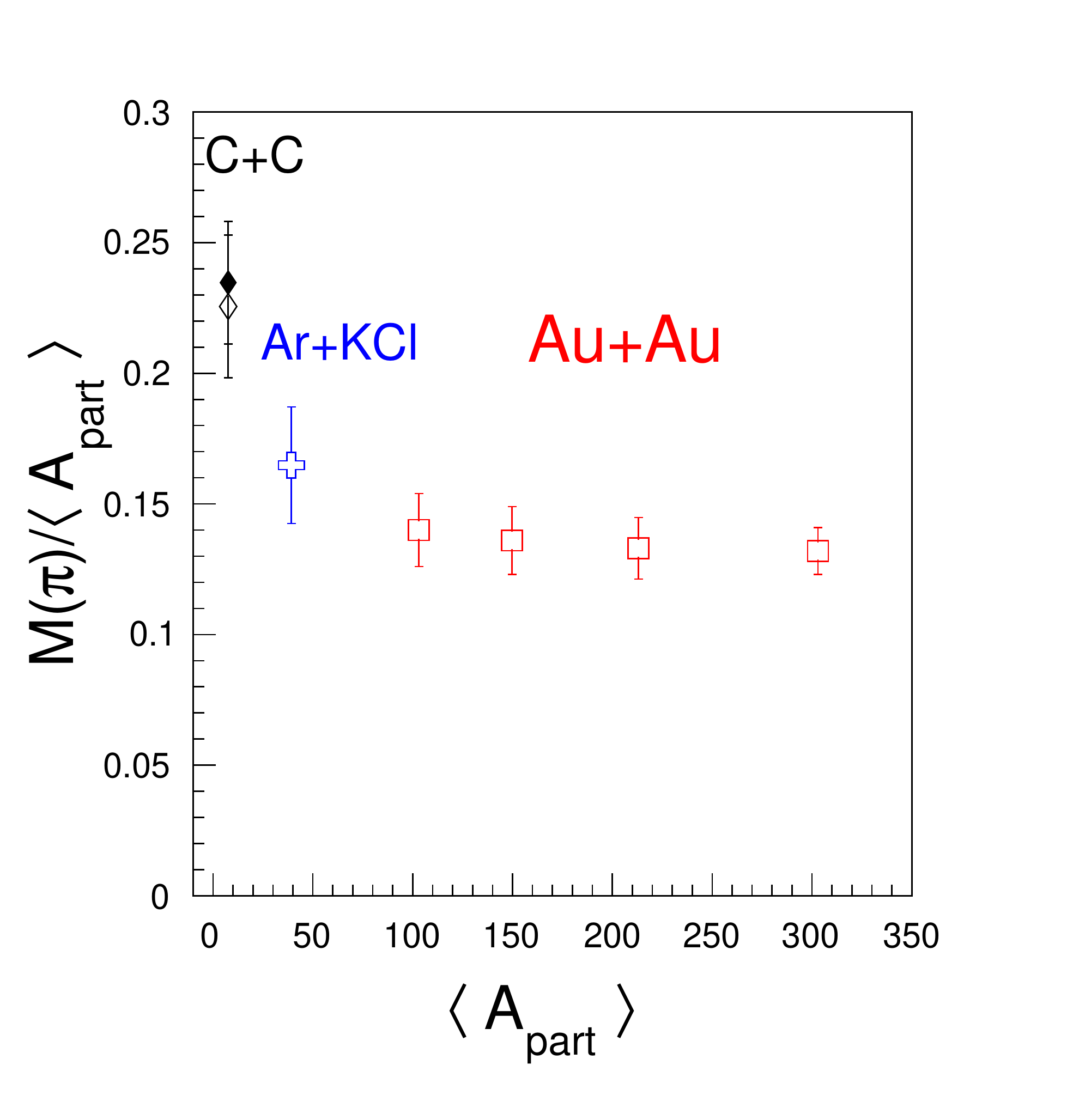}
    \caption{Pion multiplicity per participating nucleon as a function of centrality given by $\langle A_{part}\rangle$.  Here only HADES data are shown. 
    The data points for Ar+KCl~\cite{Agakishiev:2011vf} (open cross) measured at 1.76~A~GeV as well as the 
    C+C~\cite{Agakishiev:2009zv} at 1~A~GeV (closed diamond) and  at 2~A~GeV (open diamond) data points were scaled 
    to be at a beam energy of 1.23~A~GeV (for details see text).  }
    \label{fig:ApartSystems-HADES}
\end{figure}
In order to compare system size dependence of pion production measured by HADES with other experiments at slightly different energies, it is necessary to scale their results
to the same energy of 1.23~A~GeV. This kind of scaling is done to data in fig.~\ref{fig-world-Apart} using pion energy excitation function from fig.~\ref{fig-world-Ebeam}.
The centrality dependence of $M(\pi)/\langle A_{part}\rangle$ from HADES is compared to the energy-scaled FOPI data in fig.~\ref{fig-world-Apart}.  Both data sets exhibit
a similar weak variation of $M(\pi)/\langle A_{part}\rangle$ with $\langle A_{part}\rangle$ but differ significantly in absolute values.  Also shown is the reduced pion
multiplicity scaled to 1.23~A~GeV obtained by E895 \cite{Klay:2003zf} and the BEVALAC Streamer Chamber group for La+La collisions~\cite{Harris:1987md}. 
The FOPI results on pion multiplicity lie even above the La+La data in spite of the known trend that the reduced pion multiplicity increases with decreasing system size at
a given energy \cite{Gazdzicki:1995zs}.  This discrepancy might be (partly) explained by the previously mentioned different methods in the estimation of the number of participants. 
Indeed, if one uses for the FOPI data the charge balance of the reconstructed charged particles as a centrality estimator instead \cite{Reisdorf:2010aa}, 
the values for $M(\pi)/\langle A_{part}\rangle$ come closer to our data points.  

We have measured pion production also in the much lighter systems Ar+KCl~\cite{Agakishiev:2011vf,VerWest:1981dt} and C+C~\cite{Agakishiev:2009zv} at similar beam energies.
In order to examine the system-size dependence of the pion production, the yields in Ar+KCl and C+C were scaled to the beam energy of 1.23~A~GeV using the curves
displayed on fig.~\ref{fig:ApartSystems-Ebeam}. 
The scaling factors are 0.63 for Ar+KCl at 1.76~A~GeV, 1.33 for C+C at 1~A~GeV and 0.54 for C+C at 2~A~GeV. Figure~\ref{fig:ApartSystems-HADES} compares our centrality
dependent normalized pion yields from Au+Au collisions with those of the lighter systems.  While the normalized pion multiplicity varies only slightly (less than 10\%)
in collision systems with 100 participant nucleons and more it increases by up to 30\% at 40 participants and is almost a factor of two higher at 6 participants
in the light C+C system. 
\subsection{Polar angle distribution}
\label{sec3_2}
\begin{figure}
	\centering
	\includegraphics[width=1.\linewidth]{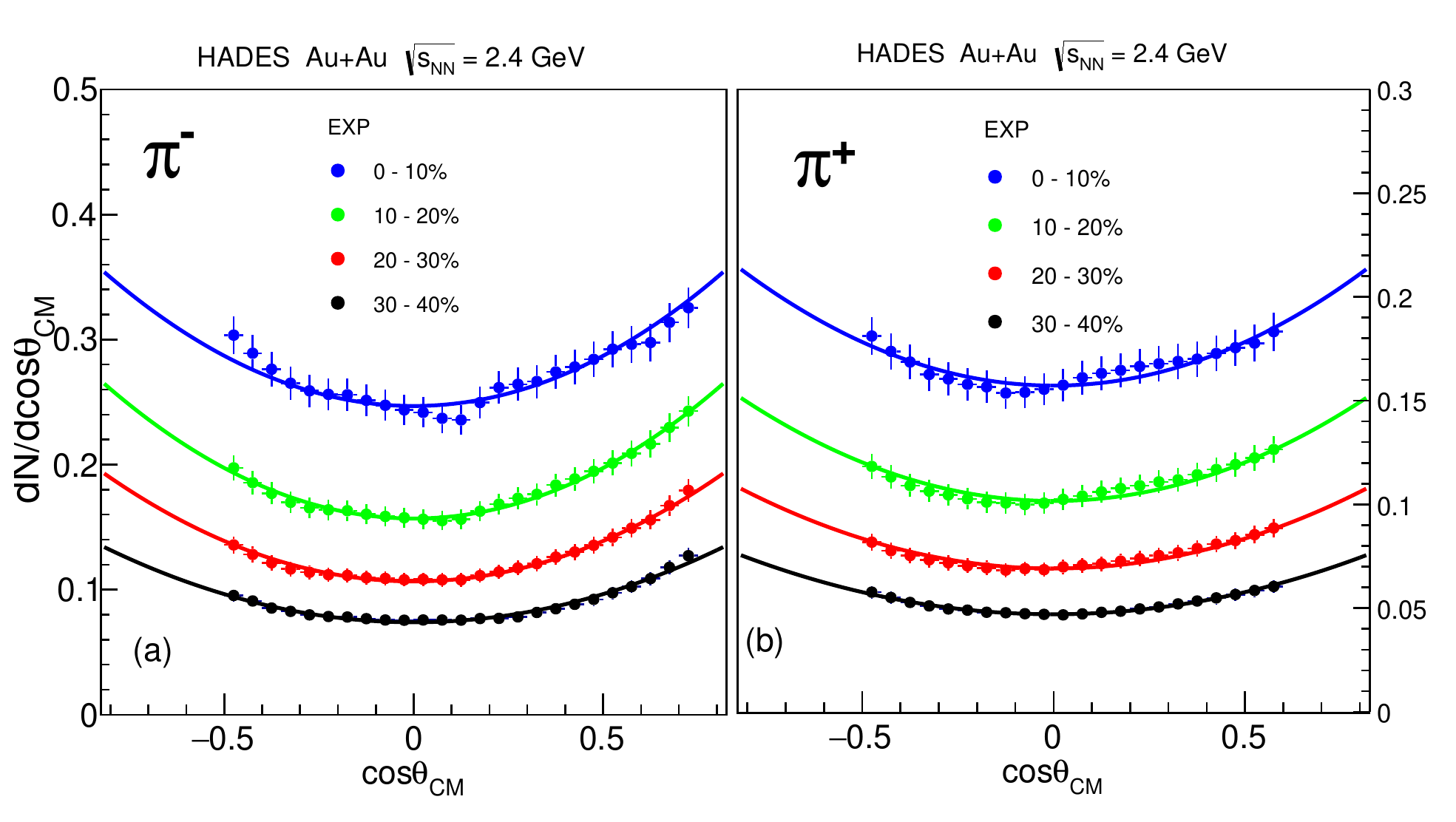}
	\caption{Center-of-mass polar angle distributions of negatively (a) and positively (b) charged $\pi$ mesons. Shown are pions with center-of-mass momenta of 120-800~MeV$/c$. The curves represent fits with the function given by eq.~\eqref{eq-angle}.}
	\label{fig-CosTheta}    
\end{figure}

So far we have parameterized the phase space of the pions by rapidity and transverse momentum or reduced transverse mass.  In the following, we span the pion phase space
using their c.m. polar angle $\theta_{cm}$ and momentum $p_{cm}$.  The polar angular distributions ($dN/d\cos\theta_{cm}$) of charged pions in heavy-ion and N+N collisions
are known to be non-isotropic with a preference for forward and backward angles.  This feature is also illustrated in fig.~\ref{fig-CosTheta} for the four centrality classes.
Fully isotropic emission would correspond to a flat distribution. 
\begin{figure}[!h]
	\centering
	\includegraphics[width=1.\linewidth]{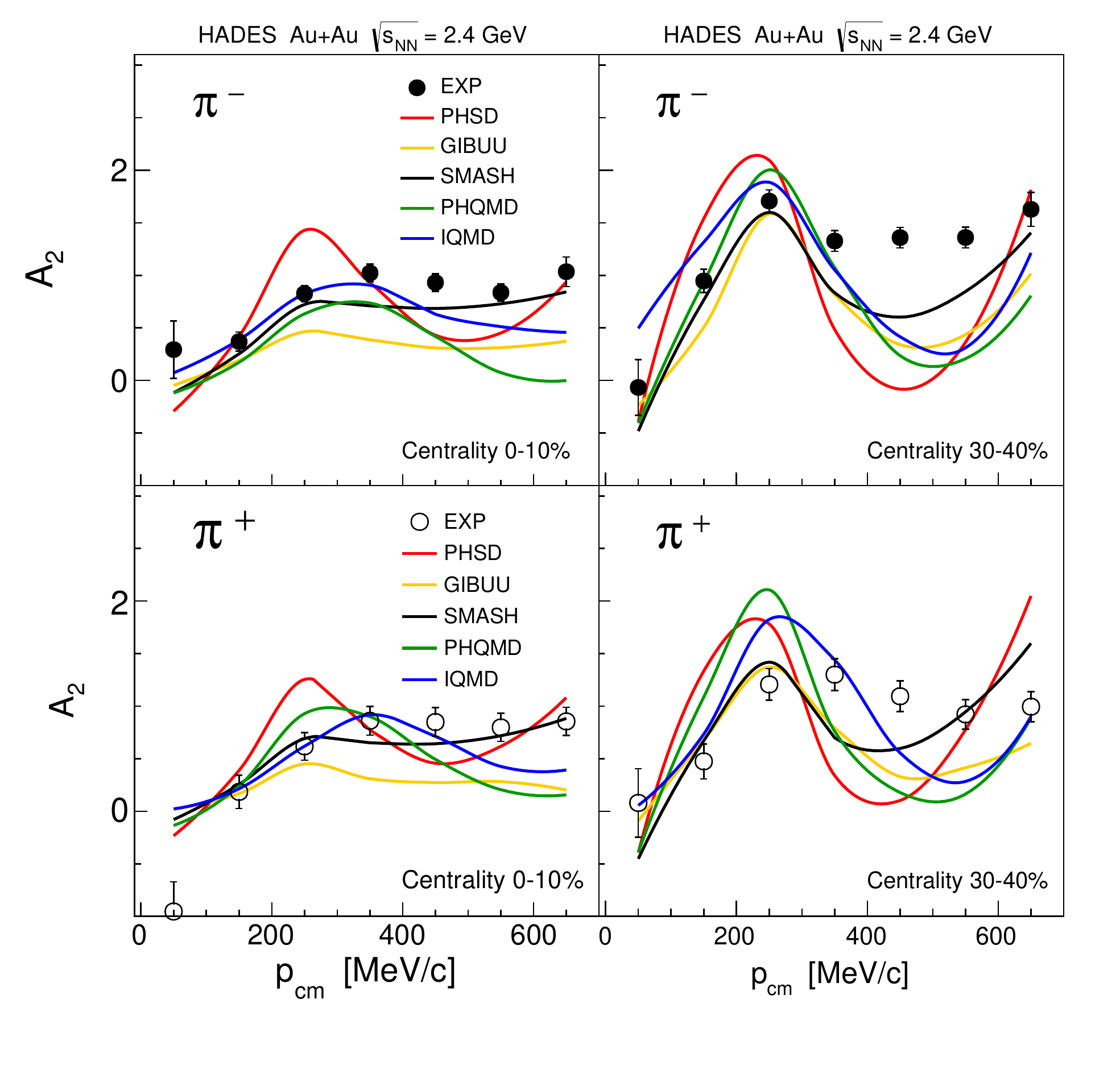}
 \caption{Dependence of the anisotropy parameter $A_{2}$ on the pion momentum in the center-of-mass system for negatively (upper row) and positively (lower row) charged pions 
 for centrality classes of 0-10~$\%$ and 30-40~$\%$. Points with error bars are the results of fits to the experimental polar angle distributions. 
 The curves represent the results of model calculations (see \secref{sec4}). Extraction of the anisotropy parameter from models is done within HADES polar angle coverage.}
\label{fig_A2_pcm}  
\end{figure}

In order to quantify the deviation from isotropy, the distributions are fitted with a quadratic function of $\cos\theta_{cm}$:

\begin{equation}
\label{eq-angle}
\frac{dN}{d(\cos\theta_{cm})} = C \,  (1 + A_{2} \, \cos^{2}\theta_{cm}),
\end{equation}
where C is a normalization factor and $A_{2}$ a parameter which quantifies the forward/backward preference of pion emission.

\begin{figure}[!h]
	\centering
	\includegraphics[width=0.4\textwidth]{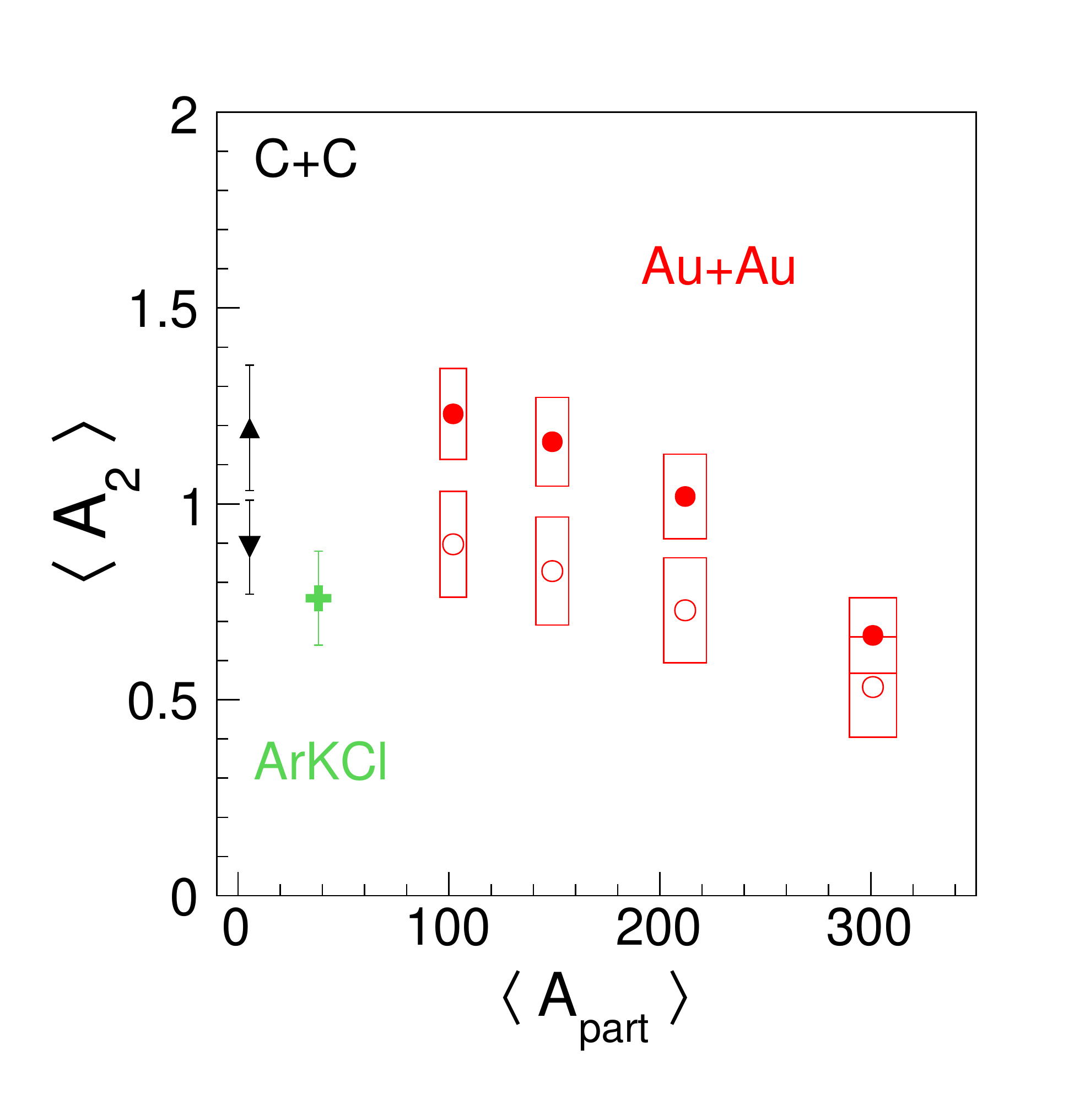}
	\caption{Mean anisotropy parameter $\langle A_{2}\rangle$ as a function of the mean number of participants $\langle A_{part}\rangle$. 
        Circles represent experimental Au+Au data (closed $\pi^{-}$, open $\pi^{+}$). 
        Also shown are earlier HADES results on $\langle A_{2}\rangle$ as average of $\pi^{-}$ and $\pi^{+}$ from C+C (triangles) and Ar+KCl (cross) collisions at similar energies. 
        The up-pointing triangle (down-pointing) stands for C+C at the beam energy of 2.0 (1.0)~A~GeV.
	}
    \label{fig-A2_Apart}       
\end{figure}

The solid curves in fig.~\ref{fig-CosTheta} are the results of the fits. The extraction of the anisotropy parameters and their comparison with models is done within the HADES acceptance.
The momentum dependence of the $A_{2}$ parameter for the most central (0-10$\%$) and the most peripheral (30-40$\%$) class is shown in fig.~\ref{fig_A2_pcm}. 
The corresponding plots for 10-20~$\%$ and 20-30~$\%$ can be seen in fig.~\ref{fig-A2-pcm_pim_pip_appendix} of the appendix (\secref{app}).
The overall trend is that in the experimental data for the most central events (0-10$\%$), $A_{2}$ is compatible with zero at low momenta ($p_{cm} \leq$ 50~MeV$/c$) 
and increases with momentum for both pion charges, saturating above 400~MeV$/c$ at $A_{2} \simeq 1.0$. In more peripheral collisions, the saturation value increases
up to $A_{2} \simeq 1.4$.  We do not observe the pion energy dependence of $A_2$ peaking around $E_{\pi}$ = 200--300 MeV as seen in Ar+KCl data at 1.8 A GeV \cite{Brockmann:1984de}.
The system-size dependence of the mean $\langle A_{2}\rangle$ (momentum-averaged over the interval 120--800~MeV$/c$) in nuclear collisions is illustrated in fig.~\ref{fig-A2_Apart}.
A weak but significant increase of $\langle A_{2}\rangle$ with decreasing system size is observed for central Au+Au collisions towards Ar+KCl and C+C collisions.  One can
define the ratio of the anisotropic to isotropic fraction R=$\langle A_2 \rangle$/(3+$\langle A_2 \rangle$).  In this way, for the most peripheral class measured by HADES (30-40$\%$),
one obtains R=0.25, and for the most central R=0.14. Using a linear extraploation to the most central collisions with $\langle A_{part} \rangle$ around 400 the anisotropic fraction
is reduced to 8$\%$.

Concerning measurements of $\langle A_{2}\rangle$ in p+p interactions at our beam energy, we found two inconsistent results.
Reference~\cite{Fickinger:1962zz} reported measurement of $dN/d\cos(\theta_{cm})$ of $\pi^+$ in $p+p\rightarrow p n \pi^+$ in a bubble chamber experiment at 2~GeV which allows to determine an 
$\langle A_{2}\rangle$ value in the range 0.8 - 1.6 which follows the trend given by our $A+A$ data in fig.~\ref{fig-A2_Apart}. In reference \cite{Brockmann:1984de},  $\langle A_{2}\rangle$ higher than 3.0 
is reported from an analysis of a counter experiment, which corresponds to R$\leq$0.5. In view of these contradicting results we refrain from presenting a data point of $\langle A_{2}\rangle$ for p+p interactions.

\section{Comparison with transport models}
\label{sec4}
\begin{figure*}[]
	\centering
        \subfigure[]{\includegraphics[width=0.35\textwidth]{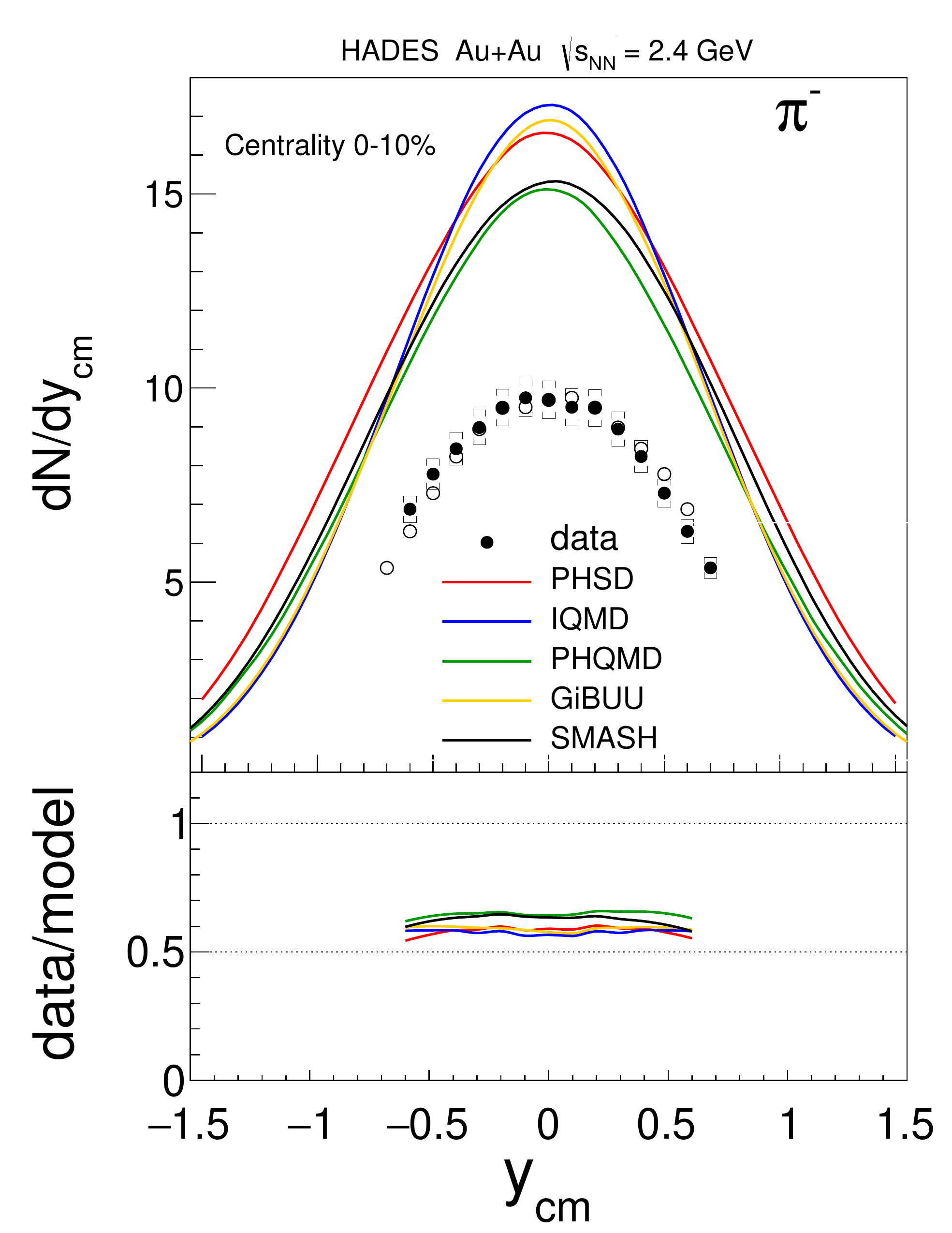}}\quad
	\subfigure[]{\includegraphics[width=0.35\textwidth]{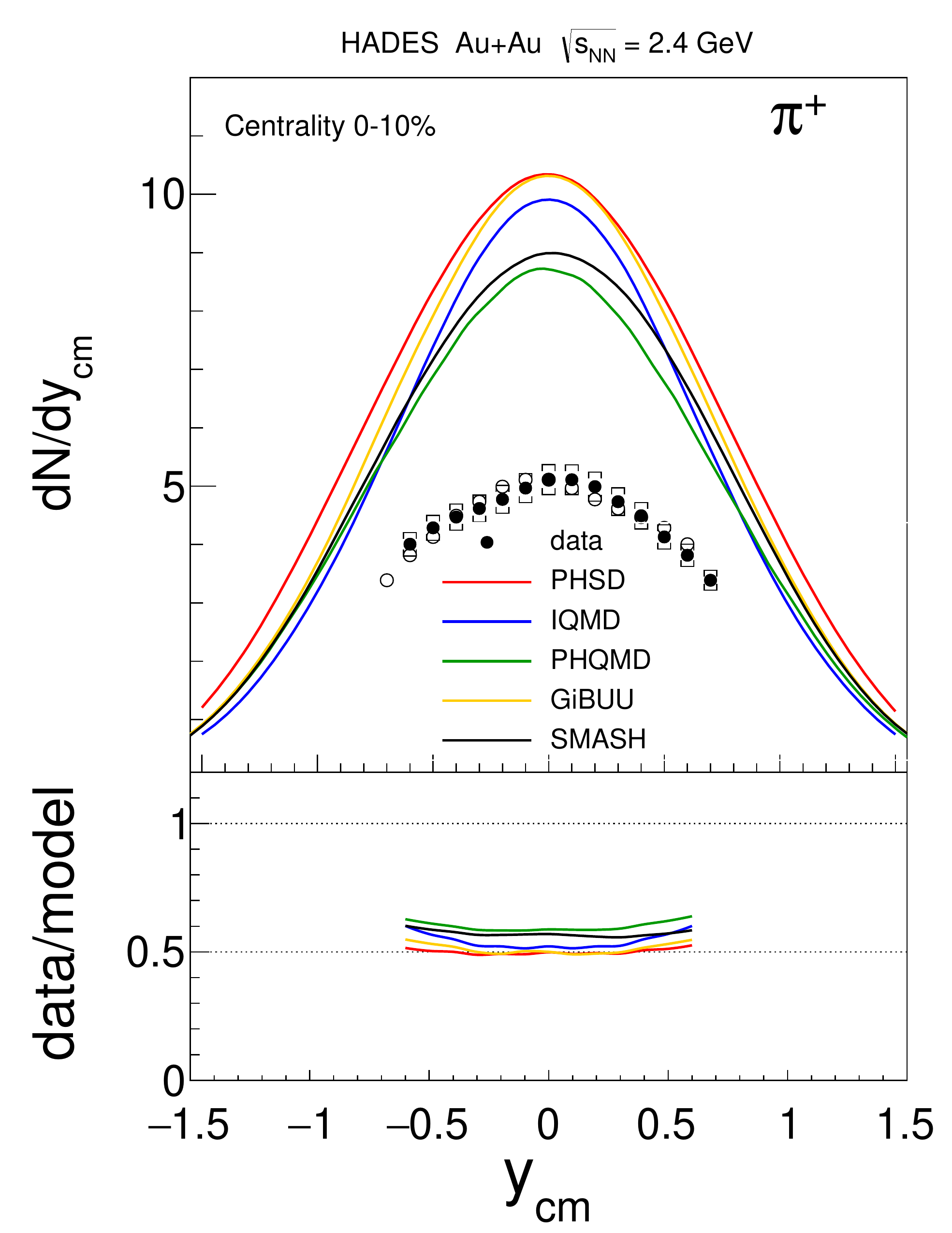}}
	\caption{Rapidity distributions of negatively (a) and positively (b) 
		charged pions. The experimental data and the results of five transport models are shown.  Full points are the measured data and open points are the data reflected
                at y$_{cm}$ = 0. The comparison is done for the 10$\%$ most central events.  The ratio of experimental over model data is displayed in the lower panels. 
		}
	\label{fig-y-theory}       
\end{figure*}

\begin{table*}[b]  
\renewcommand{\arraystretch}{1.5}
	\centering   
	\begin{tabular}{ cccccccc }
		\hline
		$\pi^{-}$  &  PHSD            & IQMD            & PHQMD          & GiBUU             & SMASH            & EXP 4$\pi$               \\ \hline
		0-10$\%$   &  30             & 28              & 26             & 27                & 28               & 17.1 $\pm$ 0.8 $\pm$ 0.9 \\
		10-20$\%$  &  20             & 19              & 19             & 20                & 19               & 12.1 $\pm$ 0.6 $\pm$ 0.6 \\
		20-30$\%$  &  14             & 13              & 13             & 14                & 13               &  8.7 $\pm$ 0.4 $\pm$ 0.4 \\
		30-40$\%$  &   9             &  8              &  9             & 10                &  9               &  6.3 $\pm$ 0.3 $\pm$ 0.3 \\ \hline
		$\alpha$   & 1.13 $\pm$ 0.01 & 1.09 $\pm$ 0.03 & 0.95 $\pm$ 0.01& 1.02 $\pm$ 0.01   & 1.03  $\pm$ 0.03 & 0.93  $\pm$ 0.01          \\ \hline\hline
		
		$\pi^{+}$  &  PHSD            & IQMD            & PHQMD          & GiBUU          & SMASH         & EXP 4$\pi$                    \\ \hline
		0-10$\%$   &  19             & 16              & 16             & 18             & 16                 &  9.3 $\pm$ 0.4 $\pm$ 0.4   \\
		10-20$\%$  &  13             & 11              & 11             & 12             & 11                 &  6.6 $\pm$ 0.3 $\pm$ 0.3   \\
		20-30$\%$  &   9             & 7               &  8             &  9             &  8                 &  4.7 $\pm$ 0.2 $\pm$ 0.2   \\
		30-40$\%$  &   6             & 4               &  6             &  6             &  5                 &  3.4 $\pm$ 0.2 $\pm$ 0.1   \\ \hline
		$\alpha$   &  1.12 $\pm$ 0.03& 1.19 $\pm$ 0.03 & 0.93 $\pm$ 0.01& 1.04 $\pm$ 0.03 & 1.03 $\pm$ 0.01   & 0.94 $\pm$ 0.01            \\ \hline
	\end{tabular}
        \vspace{0.5cm}
	\caption[]{Charged-pion multiplicities $\pi^{-}$ (top part) and $\pi^{+}$ (bottom part) in full phase space extracted from the indicated models. In the last column the 
        experimental results extrapolated to $4\pi$ are given. In the last row the values of the parameter $\alpha$ 
        are listed as obtained from the fits to the experimental data shown in fig.~\ref{fig-MultApart} and to the corresponding results of the model calculations.		
	}
	\label{tab-mult-model}
\end{table*}   
In the following the data will be compared to five state-of-the-art microspopic transport models: "Isospin Quantum Molecular Dynamics" (IQMD vi.c8)~\cite{Hartnack:1997ez},
"Parton Hadron String Dynamics" (PHSD v.4)~\cite{Cassing:1999es}, "Parton-Hadron-Quantum-Molecular Dynamics" (PHQMD v1)~\cite{Aichelin:2019tnk}, "Simulating Many Accelerated Strongly-Interacting Hadrons" (SMASH)~\cite{Petersen:2018jag,Weil:2016zrk}, 
and "Giessen Boltzmann-Uehling-Uhlenbeck" (GiBUU, release 2019)~\cite{Buss:2011mx} 
\footnote{During the preparation of the paper, we were contacted by the UrQMD authors and informed that they are revisiting and improving their code with respect to pion production at low energies. 
Therefore, we refrain from showing any results from this model.}. 
We study the differences between model predictions as observed in the rapidity, transverse momentum, and polar angle  distributions of charged pions, 
in the trends of the $A_2$ parameter as a function of $p_{cm}$ and system size, and in the predicted abundance of resonances as well. In the models, the selection of four centrality
classes is done by selecting the corresponding impact-parameter intervals (see table~\ref{tab-Apart-b}) according to the values estimated in \cite{Adamczewski-Musch:2017gjr}.
We find that all models over-predict the pion yields for all centralities by factors ranging from 1.2 to 2.1 (see  table~\ref{tab-mult-model}, fig. \ref{fig-MultApart}
and fig. \ref{fig-y-theory}).

The yields of charged pions have already been shown as a function of $\langle A_{part}\rangle$ in fig.~\ref{fig-MultApart} together with the results of the model calculations. 
Most of the model calculations give a linear or slightly stronger than linear dependence ($\alpha  \geq \ 1$), only PHQMD agrees with the significantly weaker scaling observed
in our data (see table~\ref{tab-mult-model}).

The yields and shapes of the rapidity distributions are compared in fig.~\ref{fig-y-theory}. In the lower panels the ratio of the experimental and model data quantify both
the yield excess and differences in shape.  All models tend to have slightly wider (narrower) $\pi^-$ ($\pi^+$) distributions than observed in the experiment.
\begin{figure*}[]
	\centering
	\subfigure[]{\includegraphics[width=0.35\textwidth]{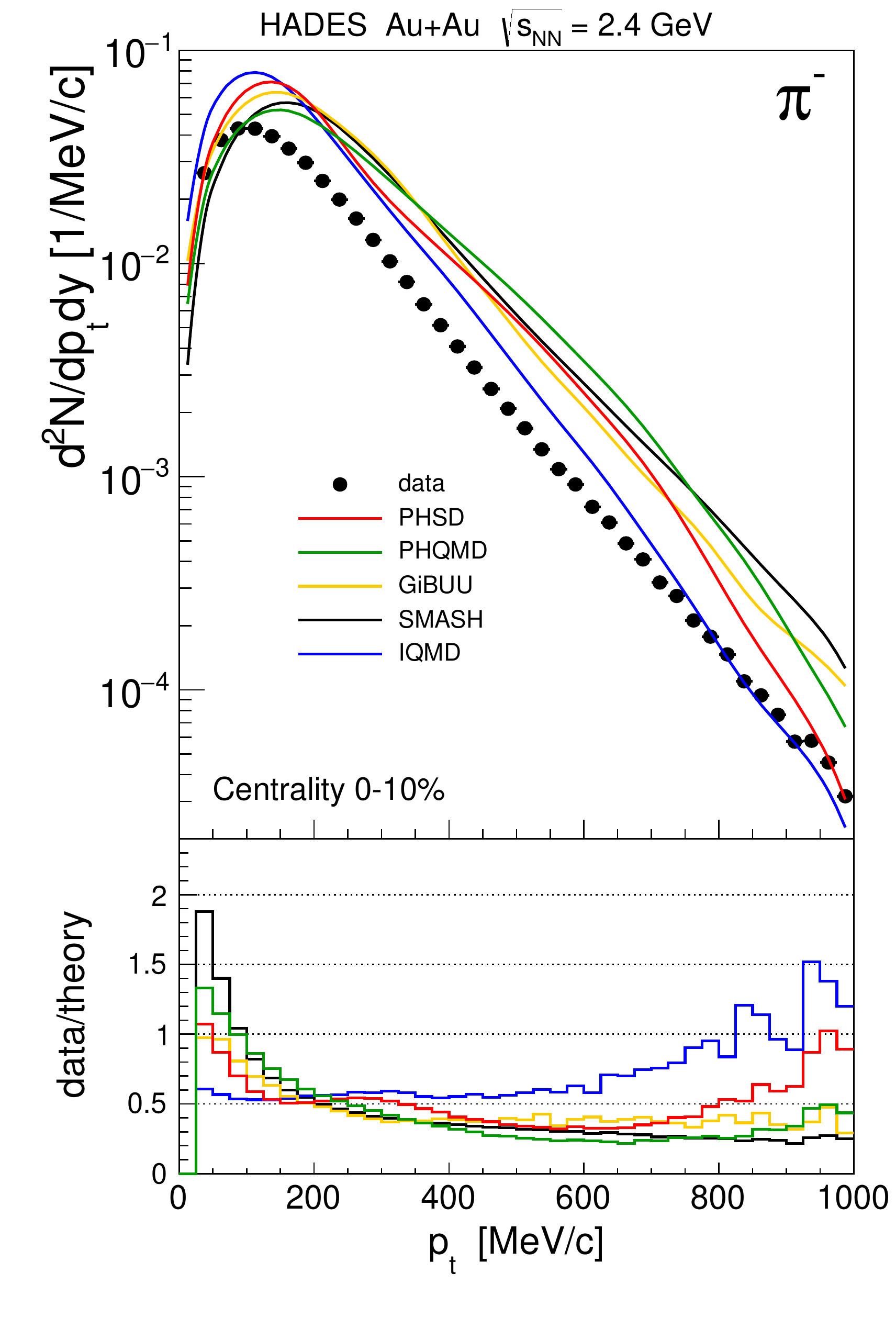}}\quad
	\subfigure[]{\includegraphics[width=0.35\textwidth]{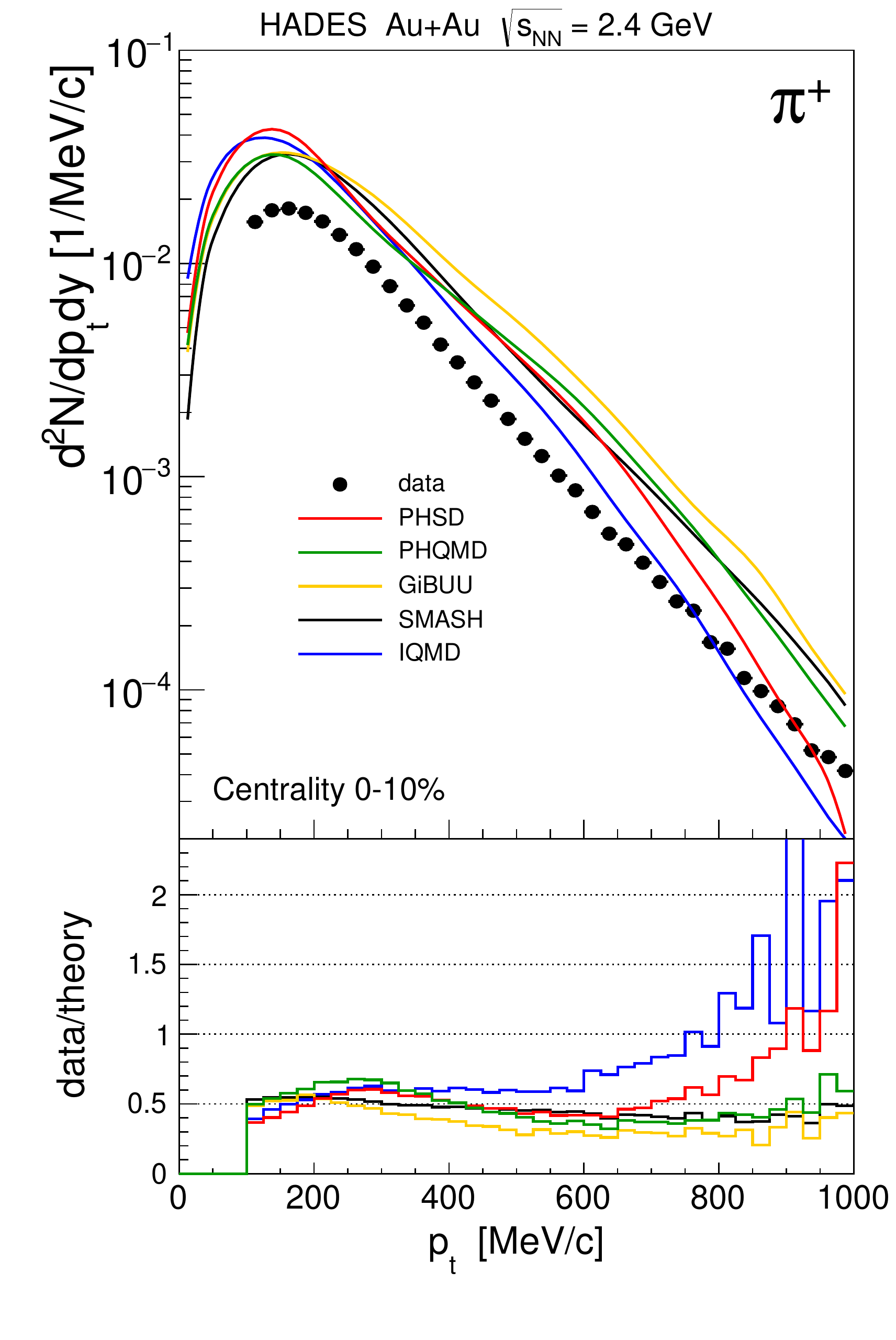}}
	\caption{The mid-rapidity (y$_{cm} \pm$ 0.05) experimental transverse-momentum distributions of
    negatively (a) and positively (b) charged pions in comparison to models for the $10\%$ most central events.
    }
	
	\label{fig-pt-theory}       
\end{figure*}
\begin{figure*}[]
	\centering
	\subfigure[]{\includegraphics[width=0.35\textwidth]{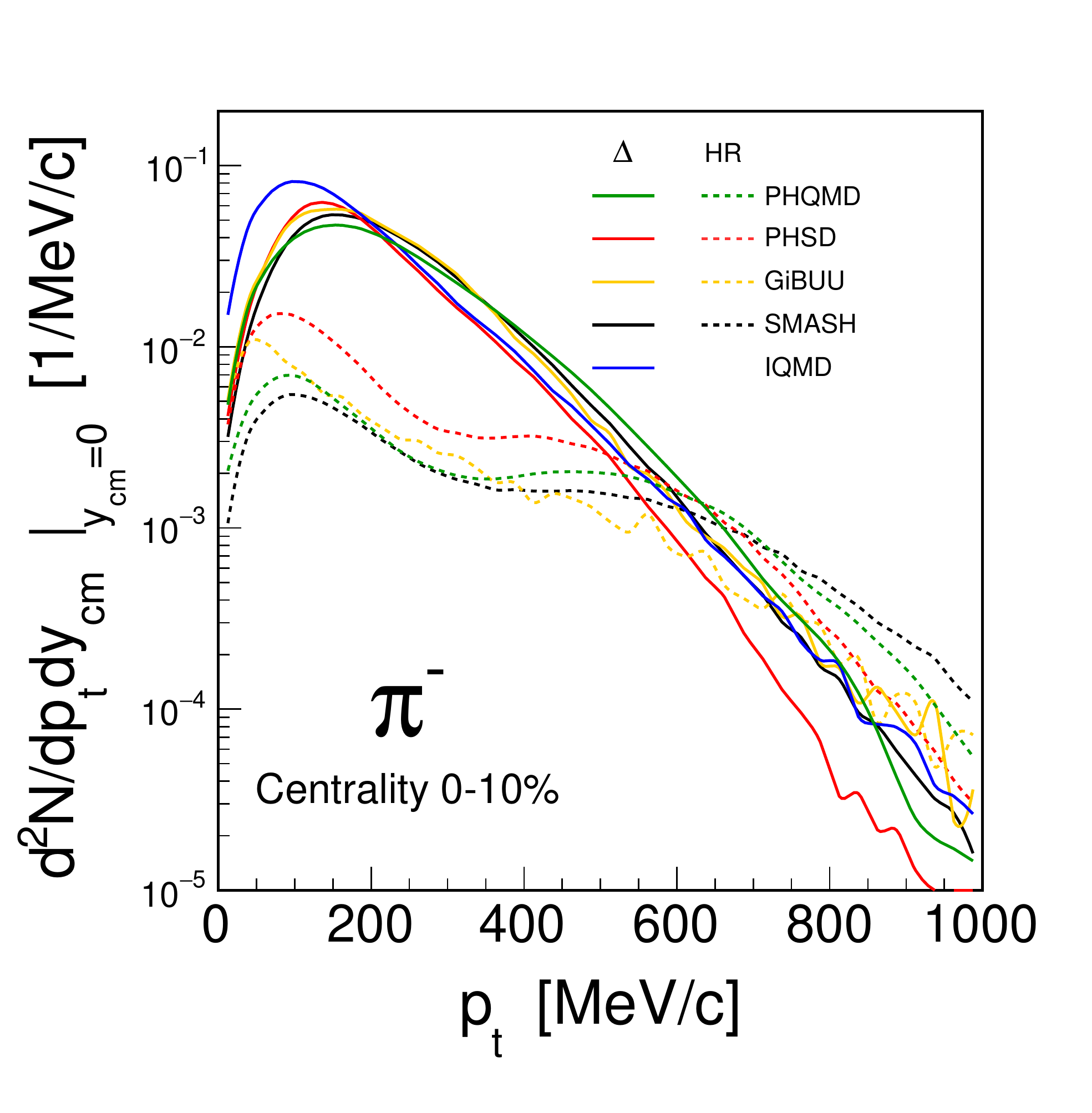}}\quad
	\subfigure[]{\includegraphics[width=0.35\textwidth]{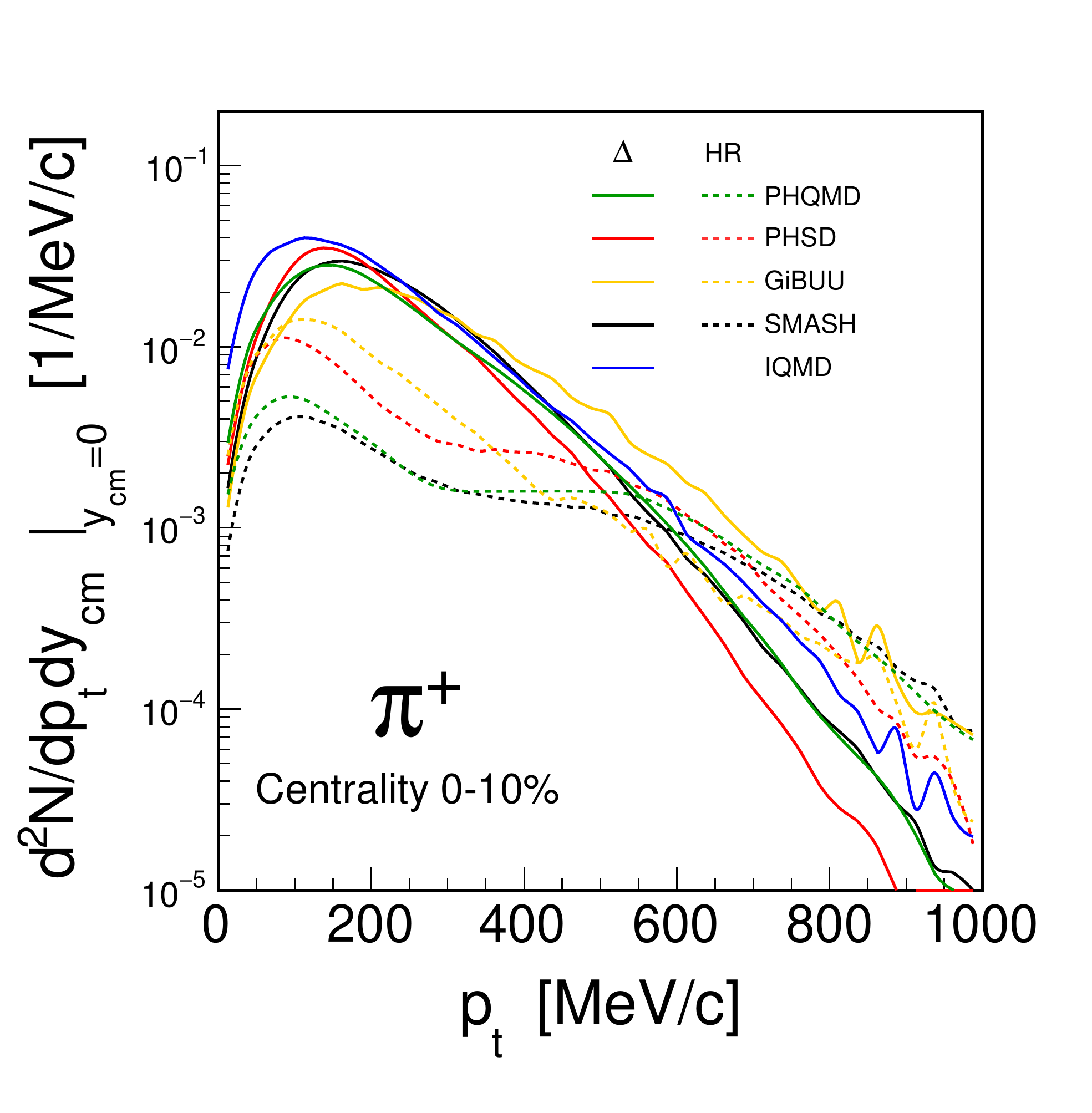}}
	\caption{Decomposition of the transport model calculations into the decay contributions from $\Delta(1232)$ (solid lines) 
        and from all higher-lying resonances (HR, dashed lines) as function of transverse momentum at mid-rapidity for $\pi^{-}$ (a) and $\pi^{+}$ (b).
	}	
	\label{fig-pt-theory-cocktail}       
\end{figure*}
The transverse-momentum distributions are compared in fig.~\ref{fig-pt-theory}.  In the case of negatively charged pions, all models besides IQMD show similar $p_{t}$ dependences.
However, the slopes are clearly steeper than observed in the data, which is in particular evident from the ratio plot (lower panels).  IQMD, on the other hand, predicts
a $p_{t}$ dependence similar to the data and thus the data/theory ratio has a rather flat dependence on $p_{t}$ up to 600~MeV$/c$.  For $\pi^{+}$, a rather flat data/theory
ratio is observed for all models.  The high $p_{t}$ region resulting from IQMD and PHQMD calculations are steeper than those of the experimental data and from the other models.
In all of the presented models, the bulk of the pion yield stems from the decays of $\Delta(1232)$ resonances.  Higher-mass states are incorporated at different levels,
as specified in table~\ref{tab-model-resonances} of the appendix (\secref{app}).  In the IQMD model, the $\Delta(1232)$ is the only source of pions and,
as discussed in \cite{Bass:1998ca}, all inelastic NN cross sections are projected onto the excitation of this particular resonance.

\begin{figure}[]
	\centering
	\includegraphics[width=0.8\linewidth]{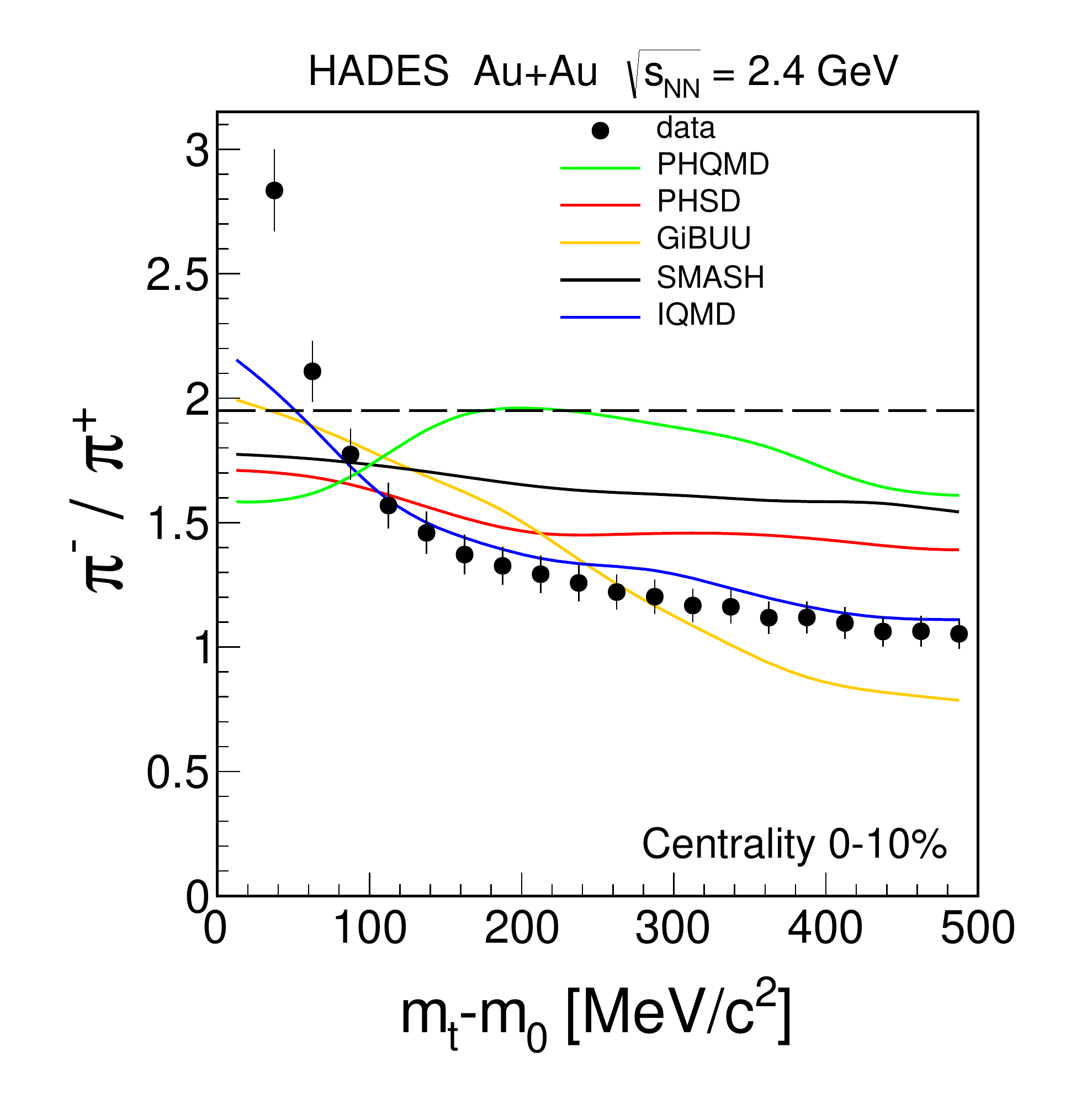}
	\caption{The measured $\pi^{-}/\pi^{+}$-ratio at mid-rapidity as a function of the reduced transverse mass in comparison to several transport model predictions. 
        The dashed line corresponds to the pion ratio of 1.94 from a simple isobar model where one assumes that all pions are produced via $\Delta$ resonances.}
	\label{fig-ratioPimPip}     
\end{figure}

Since all models overpredict the pion multiplicity significantly, one may ask whether the treatment of the resonances is at the origin of this "pion excess".
To this end, we have studied the resonance contributions to the transverse-momentum distributions of $\pi^{-}$ and $\pi^{+}$ mesons.  The result of this investigation
is shown in fig.~\ref{fig-pt-theory-cocktail}.  For both $\pi^{+}$ and $\pi^{-}$, PHQMD and SMASH have very similar shapes of the $\Delta(1232)$ component, 
but PHSD has a steeper slope than the two.  Negative pions from $\Delta$ in GiBUU are consistent with SMASH and PHQMD while positive pions have a less steep slope.
In IQMD negative pions are shifted towards lower $p_{t}$.  Overall, heavy resonances are a minor source of pions at low $p_t$ while they contribute to about 50 $\%$ to the
yield above 600~MeV$/c$.  Thus, both the pions from the $\Delta(1232)$ and those from higher-lying resonances boost up the pion yield, although in different transverse-momentum regions.
Comparing models to an analysis of exclusive channels in elementary reactions, as done e.g.\ in \cite{Agakishiev:2014wqa}, might be a promising avenue to follow in order to
scrutinize the different incorporations of higher-lying baryonic resonances in the models.  In \secref{sec2}, the mid-rapidity transverse-momentum distributions of both,
$\pi^{+}$ and $\pi^{-}$ were found to deviate significantly from the Boltzmann shape (\ref{eq-boltzmann_pt}) at low $p_{t}$.  These deviation were attributed to the Coulomb
interaction with the central positive charge distribution.  This subject is addressed again in fig.~\ref{fig-ratioPimPip} which shows the reduced transverse-mass dependence
of $\pi^{+}$ and $\pi^{-}$ ratio for data and model calculations.  The ratio remains rather constant for PHSD and SMASH, but not for PHQMD.
The small variations might be due to the different energy dependence of $\pi^{+} - p$ and $\pi^{-} - p$ inelastic cross sections, as pointed out in \cite{Bass:1995pj}. 
The experimental ratio exhibits a monotonic increase with decreasing $m_t$ and a steep rise below 100~MeV/c.  A rise at low transverse momentum is also visible in IQMD
and GiBUU, but much less pronounced.  These two transport models are the only ones which have the Coulomb interaction implemented in their codes.  The significant deviations from
data at low pt could mean that the Coulomb potential assumed in the models is smaller than in the actual collision system. 
In \secref{sec3_2} the polar angle distributions of charged pions were characterized by the parameter $A_2$.
Its dependence on $p_{cm}$ for four 10~$\%$ centrality classes was shown in fig.~\ref{fig_A2_pcm}, together with the results of the model calculations.  For the most central
(0-10~$\%$) class (upper row) the differences between data and models are moderate except for PHSD which exhibits a significant variation with $p_{cm}$.  The situation becomes
more involved if one takes the other three centrality classes into account (see fig.~\ref{fig_A2_pcm}).  Here, the structures in the data have slightly higher amplitudes. 
The models, however, develop a strong single oscillation, whose amplitude increases with decreasing centrality.
\section{Summary}
In summary, we have presented a comprehensive study of $\pi^{\pm}$ emission in Au+Au collisions at $\sqrt{s_{\rm{NN}}}=2.4$~GeV.  The data is discussed as a function of the
transverse momentum (reduced transverse mass) as well as the rapidity in four centrality classes covering the $40\%$ most central events.  We find that our results on the
pion multiplicity fit well into the overall systematic of the world data, but are lower by $2.5~\sigma$ in yield when comparing to data from a FOPI experiment performed
at a slightly lower collision energy.  The pion multiplicity per participating nucleon increases as a function of decreasing centrality and system size from 0.13 in the
$0-10~\%$ most central Au+Au collisions to 0.14 at $30-40~\%$ centrality and to 0.23 in minimum-bias C+C collisions with six participant nucleons only.
The polar angular distributions are found to be non-isotropic even for the most central event class.  The experimental data are compared to several state-of-art transport model
calculations.  All of the models substantially overestimate the absolute yields and only PHQMD is able to describe the moderate decrease of pion multiplicity per
participating nucleon as a function of increasing centrality.  While the shape of the rapidity distribution is fairly well reproduced by all models,  
the shape of the reduced transverse-mass spectra as well as the anisotropy parameter $A_2(p_{cm})$ are, however, not described satisfactorily by any. 
Decomposing the contributions to the pion spectra of the various resonances implemented in the different models, we find significant variations in both relative yield
and shape of these different pion sources.  Data on pion-induced reactions on H$_2$ and nuclear targets as well as new high-precision measurements of the collision system
Ag+Ag at $\sqrt{s_{NN}}$=2.55~GeV, taken as part of FAIR Phase0, will become available soon.  These upcoming results will be used to further extend
the world database on pion production and to constrain model calculations ever more stringently.
\vspace{1cm}
\FloatBarrier 
\begin{acknowledgement}
The HADES collaboration thanks W.~Reisdorf, N.~Herrmann, Y.~Leifels, J.~Weil, E.~Bratkovskaya, J.~Aichelin, C.~Hartnack, R.~Stock, M.~Gazdzicki, M.~Bleicher, J.~Steinheimer,
Y.~Nara, and K.~Gallmeister for elucidating discussions.
SIP JUC Cracow, Cracow (Poland), National Sience Center, 2017/25/N/ST2/00580, 2017/26/M/ST2/00600; 
TU Darmstadt, Darmstadt (Germany), VH-NG-823, DFG GRK 2128, DFG CRC-TR 211, BMBF:05P18RDFC1; 
Goethe-University, Frankfurt(Germany), HIC for FAIR (LOEWE), BMBF:06FY9100I, BMBF:05P19RFFCA, GSI F$\&$E;
Goethe-University, Frankfurt(Germany) and TU Darmstadt, Darmstadt (Germany), ExtreMe Matter Institute EMMI at GSI Darmstadt (Germany);
TU Muenchen, Garching (Germany), MLL Muenchen, DFG EClust 153, GSI TMLRG1316F, BMBF 05P15WOFCA, SFB 1258, DFG FAB898/2-2; 
NRNU MEPhI Moscow, Moscow (Russia), in framework of Russian Academic Excellence Project 02.a03.21.0005, 
Ministry of Science and Education of the Russian Federation 3.3380.2017/4.6; JLU Giessen, Giessen (Germany), 
BMBF:05P12RGGHM; IPN Orsay, Orsay Cedex (France), CNRS/IN2P3; NPI CAS, Rez, Rez (Czech Republic), 
MSMT LM2015049, OP VVV CZ.02.1.01/0.0/0.0/16 013/0001677, LTT17003.
\end{acknowledgement}
\FloatBarrier 
\section{Appendix}
\label{app}
\FloatBarrier 

\begin{figure}[!h]
 \centering
	\includegraphics[width=1.\linewidth]{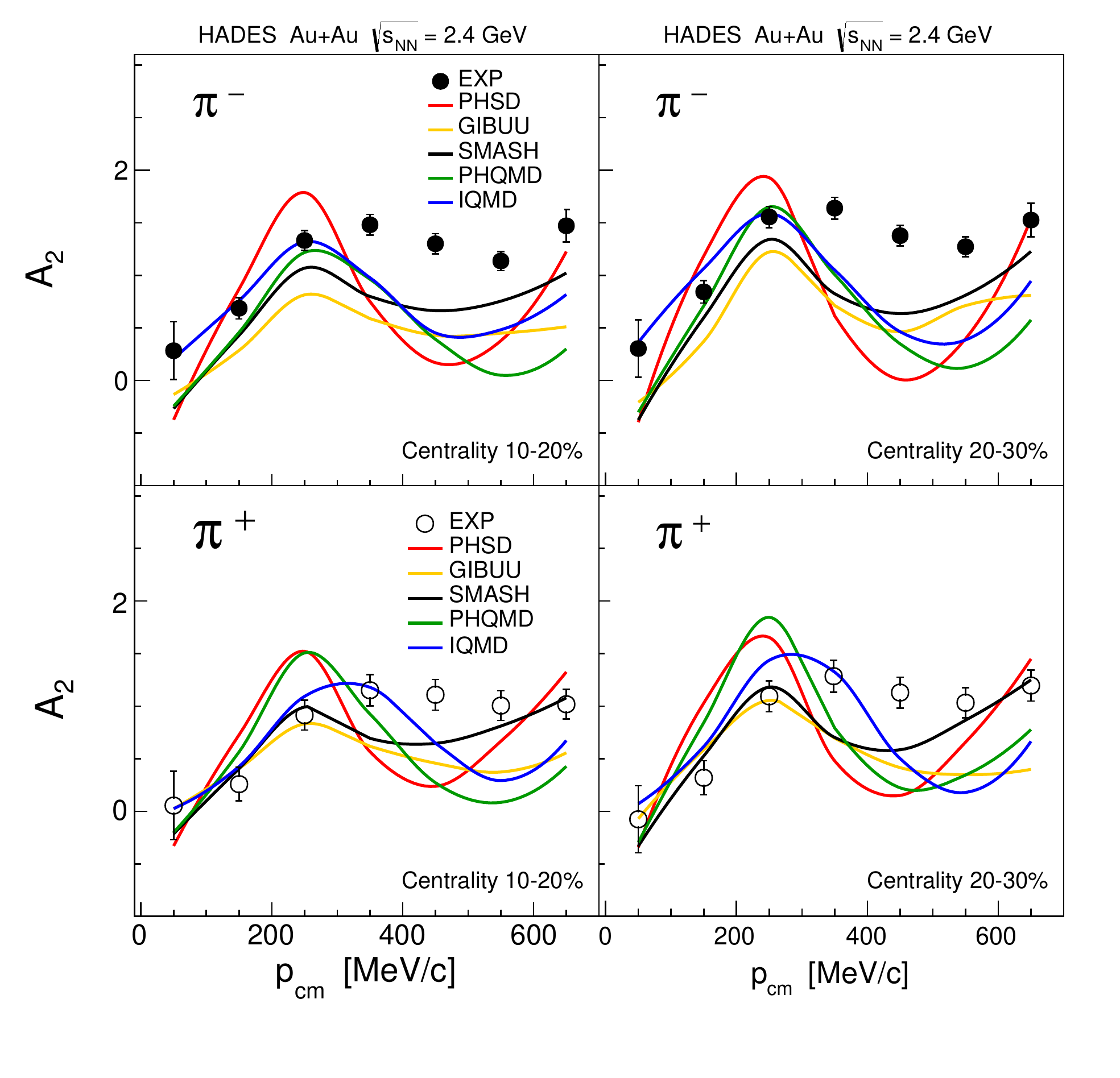}
 \caption{Dependence of the anisotropy parameter A$_{2}$ on the pion momentum in the center-of-mass system 
 for negatively charged pions (upper row) and positively (lower row)  for centrality classes of 10-20~$\%$ and 20-30~$\%$. Points with error bars are the results 
 of fits to the experimental polar angle distributions.  The curves represent the results of model calculations (see \secref{sec4}).
 Extraction of the anisotropy parameter from models is done within the HADES polar angle coverage.}
	\label{fig-A2-pcm_pim_pip_appendix}       
\end{figure}

Here supplementary information to the main text is provided.  Figure~\ref{fig-A2-pcm_pim_pip_appendix} extends fig.~\ref{fig_A2_pcm},
showing for two more centrality selections (10-20\% and 20-30\%) the dependence of the anisotropy parameter A$_{2}$ on c.m. momentum
$p_{cm}$ for negatively and positevely charged pions, respectively.
Table~\ref{tab-model-resonances} lists the baryon and meson resonances which were included in the transport calculations discussed in \secref{sec4}.

\begin{table*}[t]
\renewcommand{\arraystretch}{1.5}
\centering
\begin{tabular}{ |c|c|c|c|c||c|c|c|c|c| }
\hline
        &  Rating   & PHSD/PHQMD & SMASH     & GiBUU   &                 & Rating  & PHSD/PHQMD & SMASH     & GiBUU  \\ \hline\hline
$\eta$  &           & $\surd$   & $\surd$   & $\surd$ &  $\Delta$(1232) &  ****   & $\surd$   & $\surd$   & $\surd$  \\
$\rho$  &           & $\surd$   & $\surd$   & $\surd$ &  $\Delta$(1600) &  ****   &           &           & $\surd$  \\
$\omega$&           & $\surd$   & $\surd$   & $\surd$ &  $\Delta$(1620) &  ****   &           & $\surd$   & $\surd$  \\
N(1440) &  ****     & $\surd$   & $\surd$   & $\surd$ &  $\Delta$(1700) &  ****   &           & $\surd$   & $\surd$  \\
N(1520) &  ****     &           & $\surd$   & $\surd$ &  $\Delta$(1750) &  *      &           &           & $\surd$  \\
N(1535) &  ****     & $\surd$   & $\surd$   & $\surd$ &  $\Delta$(1900) &  ***    &           &           & $\surd$  \\
N(1650) &  ****     &           & $\surd$   & $\surd$ &  $\Delta$(1905) &  ****   &           & $\surd$   & $\surd$  \\
N(1675) &  ****     &           & $\surd$   & $\surd$ &  $\Delta$(1910) &  ****   &           & $\surd$   & $\surd$  \\
N(1680) &  ****     &           & $\surd$   & $\surd$ &  $\Delta$(1920) &   ***   &           & $\surd$   & $\surd$         \\
N(1700) &   ***     &           & $\surd$   & $\surd$ &  $\Delta$(1930) &   ***   &           & $\surd$   &  $\surd$        \\
N(1710) &  ****     &           & $\surd$   & $\surd$ &  $\Delta$(1940) &    **   &           &           &   $\surd$       \\
N(1720) &  ****     &           & $\surd$   & $\surd$ &  $\Delta$(1950) &   ****  &           & $\surd$   &   $\surd$       \\
N(1860) &  **       &           &           &         &  $\Delta$(2000) &   **    &           &           &   $\surd$       \\                      
N(1875) &  ***      &           &           &         &  $\Lambda$(1405)&  ****   & $\surd$   & $\surd$   & $\surd$  \\     
N(1880) &  ***      &           &           &         &  $\Lambda$(1520)&  ****   &           & $\surd$   & $\surd$  \\
N(1895) &  ****     &           & $\surd$   &         &  $\Lambda$(1600)&   ***   &           &           & $\surd$  \\
N(1900) &  ****     &           & $\surd$   & $\surd$ &  $\Lambda$(1670)&  ****   &           &           &   $\surd$       \\
N(1990) &    **     &           & $\surd$   & $\surd$ &  $\Lambda$(1690)&  ****   &           & $\surd$   & $\surd$  \\
N(2000) &    **     &           &           & $\surd$ &  $\Lambda$(1710)&     *   &           & $\surd$   &   \\
        &           &           &           &         &  $\Lambda$(1800)&   ***   &           & $\surd$   & $\surd$  \\
        &           &           &           &         &  $\Lambda$(1810)&   ***   &           & $\surd$   & $\surd$  \\
        &           &           &           &         &  $\Lambda$(1820)&  ****   &           &           &      $\surd$    \\
        &           &           &           &         &  $\Lambda$(1830)&  ****   &           & $\surd$   & $\surd$  \\
        &           &           &           &         &  $\Lambda$(1890)&  ****   &           &           & $\surd$  \\
        &           &           &           &         &  $\Lambda$(2000)&     *   &           &           &   \\
\hline  
\end{tabular}
\vspace{0.2cm}
\caption[]{List of baryon and meson resonances included in the transport model calculations.  Also the PDG \cite{Agashe:2014kda} rating of the particles are included (second column).
In IQMD, only the $\Delta$ is implemented.}
\label{tab-model-resonances}
\end{table*}

\FloatBarrier 

\bibliographystyle{apsrev4-2}     
\bibliography{PionPaper_v6C}
\end{document}